\documentclass[letterpaper]{article}

\usepackage[T1]{fontenc}

\usepackage{geometry}
\usepackage{setspace}

\usepackage[version=4]{mhchem} %
\usepackage[dvipsnames]{xcolor}
\usepackage{amsfonts} 
\usepackage{caption}
\usepackage{subcaption}
\usepackage{hyperref}
\usepackage{cleveref}
\usepackage{tabularx}
\usepackage{makecell}
\crefname{equation}{eq}{eqs}
\Crefname{equation}{Eq}{Eqs}

\crefname{figure}{fig}{figs}
\Crefname{figure}{Fig}{Figs}

\usepackage{tikz}
\usepackage{footmisc}
\usepackage[
    backend=bibtex,  %
    style=chem-acs,
    articletitle=true
]{biblatex}
\usepackage{enumerate}

\usepackage{bm}
\usepackage{booktabs}

\usepackage{graphicx}
\usepackage{pdfpages}
\usepackage{float}

\newfloat{scheme}{htbp}{los}
\floatname{scheme}{Scheme}
\floatname{chart}{Chart}
\newfloat{graph}{htbp}{loh}

\usetikzlibrary{shapes.geometric, arrows, positioning}

\DeclareMathOperator*{\argmin}{arg\,min}

\newcommand{\vk}{\mathbf{k}}
\newcommand{\vq}{\mathbf{q}}
\newcommand{\vG}{\mathbf{G}}
\newcommand{\vx}{\mathbf{x}}

\newcommand{\vzero}{\mathbf{0}}
\newcommand{\primedsum}{\sideset{}{'}\sum}

\newcommand{\Nocc}{N_{\text{occ}}}
\newcommand{\Nvirt}{N_{\text{virt}}}

\newcommand{\Nk}{N_{\vk}}
\newcommand{\Ng}{N_\vG}
\newcommand{\Nqg}{N_{\vq+\vG}}

\tikzstyle{process} = [rectangle, draw, text centered, minimum width=8cm, minimum height=1.2cm]
\tikzstyle{highlight} = [rectangle, draw=red, line width=2.5pt, text centered, minimum width=8cm, minimum height=1.2cm]
\tikzstyle{arrow} = [thick, ->, >=stealth]

\usepackage{authblk}

\author[1]{Stephen Jon Quiton}
\author[1]{Juan D. F. Pottecher}
\author[1]{Martin Head-Gordon*}
\author[2,3,4]{Lin Lin*}

\affil[1]{Department of Chemistry, University of California, Berkeley, California 94720, United States}
\affil[2]{Department of Mathematics, University of California, Berkeley, California 94720, United States}
\affil[3]{Applied Mathematics and Computational Research Division, Lawrence Berkeley National Laboratory, Berkeley, California 94720, United States}
\affil[4]{Computing and Mathematical Sciences, California Institute of Technology, Pasadena, California 91125, United States}

\title{Reduction of finite-size effects for second-order M{\o}ller--Plesset perturbation theory with singularity subtraction}

\date{*Email: mhg@cchem.berkeley.edu, lin@caltech.edu}

\begin{document}

\maketitle

\begin{abstract}
Second-order M{\o}ller--Plesset perturbation theory (MP2) for periodic systems suffers from finite-size errors (FSEs) that have inverse volume scaling due to the Coulomb kernel singularity in reciprocal space. This error scaling limits the routine applicability of MP2 and other methods that rely on it to real materials, requiring prohibitively dense $\vk$-point meshes for convergence toward the thermodynamic limit (TDL). We introduce MP2 singularity subtraction (MP2SS), a systematic approach that applies the singularity subtraction strategy to reduce MP2 FSEs. The method employs auxiliary functions and fitting procedures that consider both the singularities present at the origin in reciprocal space and also the discontinuities in the MP2 structure factor that arise from finite $\vk$-point sampling. We present three possible MP2SS configurations (Gaussian, exponential, and tuned) which use different combinations of decay functions and demonstrate their performance for gapped systems. Across a range of extended solids and molecular crystals at large basis sets, MP2SS reduces the prefactor in MP2 FSEs and subsequently the computational resources required to converge to the TDL. MP2SS with the Gaussian auxiliary function provides the most robust corrections. Our results establish singularity subtraction as a powerful and flexible approach for mitigating finite-size errors in periodic correlation methods and provide a foundation for extending the technique to higher-order perturbation theories and other post-Hartree-Fock (HF) methods.
\end{abstract}

\section{Introduction}
Wavefunction-based methods, such as second-order M{\o}ller--Plesset perturbation theory (MP2) and coupled-cluster (CC), have garnered recent attention in the periodic quantum chemistry literature \cite{pisaniPeriodicLocalMP22008,mcclainGaussianBasedCoupledClusterTheory2017,goldzakAccurateThermochemistryCovalent2022,gruberApplyingCoupledClusterAnsatz2018,neufeldGroundStatePropertiesMetallic2022,liangCanSpinComponentScaled2023,neugebauerAssessmentDLPNOMP2Approximations2023,zhuDLPNOMP2PeriodicSystems2025}, owing both to their effectiveness in describing certain materials and to increased compute capabilities in recent years. These \textit{ab initio} methods offer a systematically improvable suite of approaches to recover the electron correlation energy by correcting the mean-field Hartree-Fock (HF) wavefunction. In particular, MP2, which will be the focus of this work, as well as other methods that rely on MP2 (e.g. regularized and spin-scaled variants) have been shown to provide promising results across a broad range of crystals, especially those dominated by noncovalent interactions \cite{pisaniPeriodicLocalMP22008,erbaDFTLocalMP2Periodic2009,maschioLocalMP2Density2011,goldzakAccurateThermochemistryCovalent2022,chenRegularizedPerturbationTheory2025,liangCanSpinComponentScaled2023}. That being said, MP2 theory has its own deficiencies, as it is ill-suited for describing small-gap systems and results in a diverging correlation energy for metals. Furthermore, MP2 is also known to overestimate some classes of dispersion interactions such as $\pi$-stacking \cite{nguyenDivergenceManyBodyPerturbation2020}. Finally, just as in periodic HF, periodic MP2 suffers from finite-size effects, which constitute part of the deterrence against its widespread use in solid-state quantum chemistry \cite{xingStaggeredMeshMethod2021,xingUnifiedAnalysisFinitesize2024}.

Finite-size errors (FSEs) are errors in computed values that come from the $\vk$-point discretization of integrals over reciprocal space.  These errors are fundamental and specific to periodic quantum chemistry calculations, independent of basis-set incompleteness error (BSIE) and correlation errors that are also encountered in molecular calculations \cite{xingUnifiedAnalysisFinitesize2024,robinsonCondensedPhaseQuantumChemistry2025}. To eliminate FSEs, reciprocal space must be sampled with a sufficiently fine  $\vk$-point mesh in the First Brillouin Zone (FBZ). Without any corrections and assuming all orbitals $\{\psi_{i\mathbf{k}}\}$ and their energies $\{\epsilon_{i\mathbf{k}}\}$ can be computed exactly for any $\vk$, the FSE for the MP2 energy scales as $\mathcal{O}(N_{\vk}^{-1})$, where $\Nk$ is the number of $\vk$-points sampled in the FBZ. This error originates primarily from an excluded term where the Coulomb kernel is singular in reciprocal space \cite{xingUnifiedAnalysisFinitesize2024}. Several methods exist in the literature to address the FSE specifically for MP2, including the staggered mesh method \cite{xingStaggeredMeshMethod2021,xingUnifiedAnalysisFinitesize2024,xingInverseVolumeScaling2024} and cubic-spline interpolation \cite{liaoCommunicationFiniteSize2016}. The staggered mesh method, while simple in concept and implementation, incurs additional memory and compute costs owing to the calculation of the virtual space at the shifted set of $\vk$-points. Cubic-spline interpolation has also demonstrated its effectiveness for several unit cells and is easily extendable to higher order post-HF methods \cite{liaoCommunicationFiniteSize2016}, yet its success is dependent on the convergence of the MP2 structure factor with respect to $\Nk$. However, at finite $\Nk$, there are discontinuities in the structure factor for both the direct and exchange terms of MP2, even if the densities and orbital energies are themselves fully converged \cite{xingUnifiedAnalysisFinitesize2024,xingStaggeredMeshMethod2021}. These discontinuities limit the accuracy of interpolation-based methods in correcting the FSE, and thus, there is an opportunity to achieve even faster convergence to the TDL by considering the reciprocal space singularities in the MP2 energy.

In this work, we introduce the systematic treatment of the MP2 FSE with singularity subtraction, which has previously been successfully employed for exact exchange in periodic HF and hybrid density functional theory (DFT) \cite{gygiSelfconsistentHartreeFockScreenedexchange1986,quitonOptimizedAuxiliaryFunctions2025}. This strategy seeks to address the finite-size error by finding an auxiliary function that satisfies two criteria: (1) it has a singularity matching or closely resembling that of the original integrand, and (2) its integration is a closed-form expression or easily computable to an arbitrarily high accuracy. Similarly to exact-exchange singularity subtraction (ExxSS), MP2 singularity subtraction (MP2SS) involves first computing the structure factor for the MP2 correlation energy and then fitting the auxiliary function to that structure factor. For MP2, obtaining an appropriate auxiliary function requires more care, as the structure factor itself has singularities, and those singularities manifest differently in the direct and exchange terms. 

The rest of the paper is organized as follows. In \cref{sec:theory}, we review periodic MP2, classify the reciprocal-space singularities that govern its finite-size error, and introduce singularity subtraction as a numerical quadrature strategy. In \cref{sec:MP2SS}, we develop the MP2SS corrections for the direct and exchange singularities, motivate the auxiliary-function configurations considered in this work, and describe the computational procedure. In \cref{sec:results}, we assess these configurations first through exploratory calculations at a minimal basis, and then on benchmarks with extended and molecular crystals. Finally, in \cref{sec:discussion}, we summarize the performance and computational cost of MP2SS and discuss possible extensions of the method. Because MP2 diverges for metals and performs suboptimally for small-gap systems, the present study is limited to insulating systems.

\section{Theoretical Background}\label{sec:theory}
\subsection{Periodic MP2 Theory}
We work in the Bloch-orbital framework, using $i,j,k,l$ to label occupied crystalline/molecular orbitals (MOs) and $a,b,c,d$ to label virtual MOs.
We denote by $N_\mathbf{k}$ the total number of $\vk$-points in the Monkhorst-Pack mesh \cite{monkhorstSpecialPointsBrillouinzone1976}, and by $\Omega$ the unit cell, with volume $|\Omega|$. Its corresponding FBZ is $\Omega^*$, with volume $|\Omega^*|$. Throughout the paper, $\mathbf{R}\in\mathbb{L}$ denotes a real-space Bravais-lattice vector and $\mathbf{G}\in\mathbb{L}^*$ denotes a reciprocal-lattice vector. A Bloch orbital with band index $i$ and crystal momentum $\vk$ is written as $\psi_{i\mathbf{k}}(\mathbf{r})=\frac{1}{\sqrt{\Nk}} e^{\mathrm{i}\mathbf{k}\cdot\mathbf{r}}\phi_{i\mathbf{k}}(\mathbf{r})$, where $\phi_{i\mathbf{k}}(\mathbf{r})$ is the real-space periodic part of the MO. The corresponding pair density is defined as
\begin{align}
\rho_{i\mathbf{k}_{i}j\mathbf{k}_{j}}(\mathbf{r})=\phi_{i\mathbf{k}_{i}}^{*}\left(\mathbf{r}\right)\phi_{j\mathbf{k}_{j}}\left(\mathbf{r}\right):=\frac{1}{|\Omega|}\sum_{\mathbf{G}\in\mathbb{L}^{*}}\hat{\rho}_{i\mathbf{k}_{i}j\mathbf{k}_{j}}(\mathbf{G})e^{\mathrm{i}\mathbf{G}\cdot\mathbf{r}},\label{eq:pair_densities}
\end{align}
where $\hat{\rho}_{i{\mathbf{k}_{i}}j{\mathbf{k}_{j}}}(\mathbf{G})$ denotes the Fourier coefficient of the MO product.

With this definition, two-electron integrals in the Bloch orbital basis can be expressed as  
\begin{align}
\left\langle i \mathbf{k}_{i},j \mathbf{k}_{j}\mid a\mathbf{k}_{a},b\mathbf{k}_{b}\right\rangle &=\frac{1}{\left|\Omega^{s}\right|}\sideset{}{'}\sum_{\mathbf{G}\in\mathbb{L}^{*}}\hat{\rho}_{i{\mathbf{k}_{i}}a{\mathbf{k}_{a}}}(\mathbf{G})\hat{V}\left(\mathbf{q}+\mathbf{G}\right)\hat{\rho}_{j\mathbf{k}_{j}b\mathbf{k}_{b}}\left(\mathbf{G}_{\vk_i,\vk_j}^{\vk_a,\vk_b}-\mathbf{G}\right). %
\end{align}
Here $\left|\Omega^{s}\right|=N_{\mathbf{k}}|\Omega|$ denotes the volume of the periodic Born-von Karman (BvK) supercell corresponding to the discretized Brillouin zone, $\mathbf{G}_{\vk_i,\vk_j}^{\vk_a,\vk_b}:= \mathbf{k}_i + \mathbf{k}_j - \mathbf{k}_a - \mathbf{k}_b \in \mathbb{L}^*$ by crystal momentum conservation,  $\mathbf{q}=\mathbf{k}_a-\mathbf{k}_i$ is the momentum-transfer vector, $\sum'$ indicates that the (singular) term where 
$\mathbf{q}+\mathbf{G}=\mathbf{0}$ is excluded, and $\hat{V}(\vq+\vG)=4\pi/|\vq+\vG|^2$ is the Coulomb kernel in reciprocal space. If $\mathbf{k}_i, \mathbf{k}_a$ belong to a grid in the FBZ denoted by $\mathcal{K}$, then $\mathbf{q}$ belongs to an origin-centered grid denoted by $\mathcal{K}_\vq$. The tessellation of  $\mathcal{K}_\vq$ across the whole domain of $\mathbb{R}^3$ is denoted by $\mathcal{K}_\vq + \mathbb{L}^* := \{ \mathbf{q} + \mathbf{G}\mid\mathbf{q} \in \mathcal{K}_\vq,  \mathbf{G} \in \mathbb{L}^*\}$.
 
The restricted-orbital closed shell MP2 correlation energy at the finite $\Nk$ can first be expressed as a triple quadrature over the FBZ of the regular MP2 expression
\begin{align}
E_{\mathrm{MP2}}(\Nk)
=
\frac{1}{\Nk}
\sum_{ijab}
\sum_{\mathbf{k}_i \mathbf{k}_j \mathbf{k}_a \in \mathcal{K}}
\frac{
2\langle i\mathbf{k}_i, j\mathbf{k}_j | a\mathbf{k}_a, b\mathbf{k}_b \rangle
-
\langle i\mathbf{k}_i, j\mathbf{k}_j | b\mathbf{k}_b, a\mathbf{k}_a \rangle
}{
\varepsilon^{a\mathbf{k}_a, b\mathbf{k}_b}_{i\mathbf{k}_i, j\mathbf{k}_j}
}
\langle a\mathbf{k}_a, b\mathbf{k}_b | i\mathbf{k}_i, j\mathbf{k}_j \rangle,\label{eq:mp2_corr_energy_nk}
\end{align}
where $\mathbf{k}_b$ is determined by crystal momentum conservation, and the energy denominator is 
\begin{equation}
\varepsilon^{a\mathbf{k}_a, b\mathbf{k}_b}_{i\mathbf{k}_i, j\mathbf{k}_j}
:=
\varepsilon_{i\mathbf{k}_i}
+
\varepsilon_{j\mathbf{k}_j}
-
\varepsilon_{a(\mathbf{k}_i+\mathbf{q})}
-
\varepsilon_{b(\mathbf{k}_j-\mathbf{q})}.
\end{equation}
The TDL is achieved as the mesh $\mathcal{K}$ becomes dense in $\Omega^*$ and each Riemann sum $\frac{1}{\Nk}\sum_{\mathbf{k}\in \mathcal{K}}$ converges to the corresponding integral $\frac{1}{|\Omega^*|}\int_{\Omega^*} d\mathbf{k}$. 

Our discussion on FSEs for MP2 focuses on the part introduced by the quadratures involving the Coulomb-kernel singularity. Previous analyses \cite{xingUnifiedAnalysisFinitesize2024,xingStaggeredMeshMethod2021} show that, for insulating systems and in the absence of additional non-smoothness from the orbital energies or amplitudes, this is the leading contribution to the inverse-volume scaling of the error. Additional errors arising from the orbitals $\{\psi_{i\vk}\}$, the orbital energies $\{\epsilon_{i\vk}\}$, and basis-set incompleteness, while important, are not considered in this work.

\subsection{Classification of the MP2 Singularities} \label{subsec:singularities_classification}
The direct term of the MP2 energy (the first term in the numerator of \eqref{eq:mp2_corr_energy_nk}) at the TDL can be expressed as an integral over the FBZ.

\begin{align}
E_{\mathrm{MP2,d}}^{\mathrm{TDL}}
=
\frac{1}{|\Omega^*|}
\int_{\Omega^*} d\mathbf{q}
\sum_{\mathbf{G},\mathbf{G}'\in\mathbb{L}^*}
\frac{
S_{\mathrm{d}}(\mathbf{q},\mathbf{G},\mathbf{G}')
}{
|\mathbf{q}+\mathbf{G}|^2
|\mathbf{q}+\mathbf{G}'|^2
},\label{eq:emp2_d_GGprime}
\end{align}
where $S_{\mathrm{d}}(\mathbf{q},\mathbf{G},\mathbf{G}')$ is the direct term structure factor, defined as

\begin{align} \label{eq:unfolded-direct-structure-factor-full-defition}
S_{\mathrm{d}}(\mathbf{q},\mathbf{G},\mathbf{G}')
&=
\frac{2}{|\Omega^*|^2}
\int_{\Omega^*} d\mathbf{k}_i
\int_{\Omega^*} d\mathbf{k}_j
\sum_{ijab}\frac{
r_{ijab}(\mathbf{k}_i,\mathbf{k}_j,\mathbf{q}+\mathbf{G})
\, r_{ijab}(\mathbf{k}_i,\mathbf{k}_j,\mathbf{q}+\mathbf{G}')^{*}}{\varepsilon^{a\mathbf{k}_a, b\mathbf{k}_b}_{i\mathbf{k}_i, j\mathbf{k}_j}
},
\end{align} 
and $r_{ijab}(\mathbf{k}_i,\mathbf{k}_j,\mathbf{q}+\mathbf{G}) = \frac{4\pi}{|\Omega|}\hat{\rho}_{i{\mathbf{k}_{i}}a{(\vk_i+\vq)}}(\mathbf{G})\hat{\rho}_{j\mathbf{k}_{j}b(\vk_j-\vq)}\left(-\mathbf{G}\right)$ is the product of the Fourier transforms of two MO products. Note that in this definition, which utilizes $\vq$ instead of $\vk_a$ and $\vk_b$, crystal momentum conservation holds since  $\mathbf{G}_{\vk_i,\vk_j}^{\vk_a,\vk_b}=\mathbf{G}_{\vk_i,\vk_j}^{\vk_i+\vq,\vk_j-\vq}=\vzero$.
Alternatively, with $\Delta \vG = \vG'-\vG$ and taking advantage of the symmetry of the direct structure factor,
$S_{\mathrm{d}}(\mathbf{q} + \mathbf{G}, \vzero, \mathbf{G}' - \mathbf{G})=S_{\mathrm{d}}(\mathbf{q}, \mathbf{G}, \mathbf{G}')$,
we can rewrite the MP2 direct energy as
\begin{align}
E_{\mathrm{MP2,d}}^{\mathrm{TDL}}
&=
\frac{1}{|\Omega^*|}
\int_{\mathbb{R}^3} d\mathbf{q}
\sum_{\Delta \mathbf{G}\in\mathbb{L}^*}
\frac{
S_{\mathrm{d}}(\mathbf{q},\vzero, \Delta \mathbf{G})
}{
|\mathbf{q}|^2 \, |\mathbf{q}+\Delta \mathbf{G}|^2
}
=
\frac{1}{|\Omega^*|}
\int_{\mathbb{R}^3} d\mathbf{q}
\left[
\frac{
S_{\mathrm{d}}(\mathbf{q},\vzero, \vzero)
}{
|\mathbf{q}|^4
}
+
\sum_{\Delta \mathbf{G}\neq \vzero}
\frac{
S_{\mathrm{d}}(\mathbf{q},\vzero, \Delta \mathbf{G})
}{
|\mathbf{q}|^2 \, |\mathbf{q}+\Delta \mathbf{G}|^2
}
\right]
\label{eq:mp2_direct_tdl_deltaG_unfolded}
\end{align}
This reduction from an FBZ integral with two reciprocal-lattice sums to an integral over $\mathbb{R}^3$ turns $N_\vG^2$ singularity-subtraction problems over $\Omega^*$ into $N_\vG$ problems over all of reciprocal space. From \cref{eq:mp2_direct_tdl_deltaG_unfolded} we observe two types of singularities. The first is a fourth-order (quartic) singularity at $\Delta\vG=\vzero$ as $\vq\rightarrow\vzero$, represented by the first term. The second is a collection of second-order (quadratic) singularities when $\Delta\vG\neq\vzero$, occurring as $\vq\rightarrow\vzero$ or $\vq\rightarrow-\Delta\vG$. Under the same smoothness and gap assumptions stated above, both types of singularities produce $\mathcal{O}(\Nk^{-1})$ contributions to the MP2 FSE.

Analogously, the exchange part of the MP2 correlation energy (i.e. arising from the second term in the numerator of \cref{eq:mp2_corr_energy_nk}) at the TDL can be expressed as a double integral over the FBZ:
\begin{align}
E_{\mathrm{MP2,x}}^{\mathrm{TDL}}
=
\frac{1}{|\Omega^*|^2}
\int_{\Omega^*} d\mathbf{q}_1
\int_{\Omega^*} d\mathbf{q}_2
\sum_{\mathbf{G},\mathbf{G}'\in\mathbb{L}^*}
\frac{
S_{\mathrm{x}}(\mathbf{q}_1,\mathbf{q}_2,\mathbf{G},\mathbf{G}')
}{
|\mathbf{q}_1+\mathbf{G}|^2
|\mathbf{q}_2+\mathbf{G}'|^2
},\label{eq:mp2_exchange_tdl_unfolded_1}
\end{align}
where the unfolded exchange structure factor is defined as 
\begin{align} \label{eq:unfolded-exchange-structure-factor-full-definition}
S_{\mathrm{x}}(\mathbf{q}_1,\mathbf{q}_2,\mathbf{G},\mathbf{G}')
&=
-\frac{1}{|\Omega^*|}
\int_{\Omega^*} d\mathbf{k}_i
\sum_{ijab}\frac{
r_{ijba}
\bigl(
\mathbf{k}_i,
\mathbf{k}_i+\mathbf{q}_1+\mathbf{q}_2,
\mathbf{q}_1+\mathbf{G}
\bigr) %
r_{abij}
\bigl(
\mathbf{k}_i+\mathbf{q}_2,
\mathbf{k}_i+\mathbf{q}_1,
-\mathbf{q}_2-\mathbf{G}'
\bigr)}{
\varepsilon^{a(\vk_i+\vq_2), b(\vk_i+\vq_1)}_{i\mathbf{k}_i, j(\mathbf{k}_i + \mathbf{q}_1 + \mathbf{q}_2)}}.
\end{align}
Above, we use the convention
$
\mathbf{q}_2 = \mathbf{k}_a - \mathbf{k}_i,
\mathbf{q}_1 = \mathbf{k}_b - \mathbf{k}_i,$
and therefore by crystal momentum conservation,
$
\mathbf{k}_j
=
\mathbf{k}_i + \mathbf{q}_1 + \mathbf{q}_2.
$
Note that because the sums over the reciprocal lattice are taken over all of $\mathbb{L}^*$, the exchange structure factor obeys
$
S_{\mathrm{x}}(\mathbf{q}_1,\mathbf{q}_2,\mathbf{G},\mathbf{G}')
=
S_{\mathrm{x}}(\mathbf{q}_1+\mathbf{G},\mathbf{q}_2+\mathbf{G}',\mathbf{0},\mathbf{0}).
$
In other words, the structure factor can be re-expressed to depend only on two independent wavevectors. Consequently, eq \eqref{eq:mp2_exchange_tdl_unfolded_1} can be rewritten as a double integral over all of reciprocal space:
\begin{align}
E_{\mathrm{MP2,x}}^{\mathrm{TDL}}
=
\frac{1}{|\Omega^*|^2}
\int_{\mathbb{R}^3} d\mathbf{q}_1
\int_{\mathbb{R}^3} d\mathbf{q}_2
\frac{S_{\mathrm{x}}(\mathbf{q}_1,\mathbf{q}_2,\vzero,\vzero)}{|\mathbf{q}_1|^2 |\mathbf{q}_2|^2},\label{eq:mp2_exchange_tdl_folded}
\end{align}
The finite-size error in the exchange term when discretizing over $\vq_1$ and $\vq_2$ comes from the singularities at $\vq_1=\vq_2=\vzero$. Under the same regularity assumptions used for the direct term, this contribution is also $\mathcal{O}(\Nk^{-1})$ \cite{xingStaggeredMeshMethod2021,xingUnifiedAnalysisFinitesize2024}.

\subsection{Numerical Quadrature and Singularity Subtraction (SS) }\label{sec:successful-quadrature}
Uniform quadrature is the simplest numerical integration technique, in which the integrand is discretized into equally-sized pieces over a uniform grid, and the integral is approximated by summing these pieces with equal weighting. Most periodic electronic structure codes are implicitly using uniform quadrature when they use uniform k-grids to sample the FBZ.

There are three aspects of an integral that determine the performance of an approximation by uniform quadrature: (1) the domain of integration, (2) the periodicity of the integrand, and (3) the regularity of the integrand (e.g., how many continuous derivatives it has). For our purposes, a uniform quadrature on a regular grid is effective in the following two settings \cite{trefethenExponentiallyConvergentTrapezoidal2014}: 

\begin{enumerate}[(A)]
    \item If the domain of integration is the full space $\mathbb{R}^d$, then convergence can be rapid provided the integrand is sufficiently smooth and decays sufficiently fast at infinity.
    \item If the domain of integration is a $d$-dimensional parallelepiped in $\mathbb{R}^d$, then convergence can be rapid provided the integrand is periodic on that domain and sufficiently smooth.
\end{enumerate}
In both cases, higher regularity improves the asymptotic rate, but the precise rate depends on the available smoothness; with only finitely many derivatives one generally expects algebraic convergence, whereas exponential convergence requires stronger analyticity assumptions. Thus, one should avoid employing uniform quadrature on integrands that are not periodic on a finite domain of integration or on integrands that contain singularities or cusps, since these features degrade the regularity needed for rapid convergence and can dominate the discretization error.

Singularity subtraction (SS)\cite{aliabadiTaylorExpansionsSingular1985,davisMethodsNumericalIntegration1975} is a technique that reduces quadrature errors of singular integrands by adding and subtracting an auxiliary function that reproduces the leading local singular behavior of the integrand \cite{gygiSelfconsistentHartreeFockScreenedexchange1986,massiddaHartreeFockLAPWApproach1993,broqvistHybridfunctionalCalculationsPlanewave2009,quitonOptimizedAuxiliaryFunctions2025}. The goal is not to remove the singular contribution from the problem, but to isolate it into a term that can be integrated analytically or to very high accuracy, leaving behind a smoother remainder that is better suited to uniform quadrature. For example, if one is trying to compute the integral over $\mathbb{R}^d$ of a function $f(\vx)$ that has an isolated integrable singularity at $\vx=\vzero$ with uniform quadrature (on an origin-centered grid), then the naive approach would be to simply exclude the singular point and apply quadrature to the rest of the domain, i.e.,
\begin{align}
    \int_{\mathbb{R}^d}f(\vx)\mathrm{d}\vx\approx \primedsum_{i=1}^M  w(\vx_i) f(\vx_i),\label{eq:regular_quadrature}
\end{align}
where the primed sum means the singular point is excluded, and $w(\vx_i)$ is the quadrature weight at the node $i$. 
The singularity and the excluded quadrature point hinder convergence towards the true value of the integral as the number of quadrature points increases.

The SS method accelerates this convergence by introducing an auxiliary function $g(\vx)$, resulting in the following splitting of the integrand %
\begin{align}\label{eq:singularity-subtraction-general}
\int_{\mathbb{R}^d}f(\vx)\mathrm{d}\vx=\int_{\mathbb{R}^d}(f(\vx)-g(\vx))\mathrm{d}\vx + \int_{\mathbb{R}^d}g(\vx)\mathrm{d}\vx \approx \primedsum_i  w(\vx_i)(f(\vx_i)-g(\vx_i))+\int_{\mathbb{R}^d}g(\vx)\mathrm{d}\vx.
\end{align}
Here, $g(\vx)$ should be chosen so that the second term (the integral involving $g$) has a closed-form expression or is easily computable, and $g(\vx)$ should reproduce the leading local asymptotic behavior of $f(\vx)$ near $\vx = \vzero$. This improves the asymptotic convergence rate for the quadrature of $f-g$. If, in addition, $g(\vx)$ also matches $f(\vx)$ well away from the singularity, then the quadrature term $\sum_i  w(\vx_i)(f(\vx_i)-g(\vx_i))$ is also small, and this reduces the error prefactor. Just as in the case with HF exact exchange, good performance in the pre-asymptotic regime is critical to the success of SS schemes when applied to the MP2 correlation energy, and, as will be discussed at length in this work, a careful selection of auxiliary function is required to systematically treat the singularities identified above. The reformulation of the MP2 correlation energy in \cref{eq:mp2_direct_tdl_deltaG_unfolded,eq:mp2_exchange_tdl_folded} allows us to apply SS in the setting (A) where the domain of integration is $\mathbb{R}^d$, which will be discussed in more detail next.

\section{Singularity Subtraction Applied to Periodic MP2} \label{sec:MP2SS}

We now apply SS to the MP2 correlation energy. Unlike HF exact exchange, MP2 has multiple types of singularities in reciprocal space as expressed in \cref{eq:mp2_direct_tdl_deltaG_unfolded,eq:mp2_exchange_tdl_folded}. Each singularity type needs special care in terms of auxiliary function design and fitting strategy. The exchange term singularities, while still important to correct, are not expected to contribute as much to the FSE as the fourth and second-order singularities in the direct term, which are of comparable magnitude to each other in their respective FSE contributions. 

\paragraph{Fourth-order singularity in the direct term.}
Applying the idea of SS to this term yields the following contribution to the approximate MP2 energy.
\begin{align}
    E^{\textrm{SS}}_\textrm{MP2} \leftarrow \frac{1}{\Nk}\primedsum_{\mathbf{q} \in \mathcal{K}_\vq + \mathbb{L}^*}\frac{S_\mathrm{d}(\mathbf{q},\vzero,\vzero)-A_\mathrm{d}(\mathbf{q},\mathbf{0};\bm{\alpha})}{|\mathbf{q}|^4} + \frac{1}{|\Omega^*|}\int_{\mathbb{R}^3} d\mathbf{q} \frac{A_\mathrm{d}(\mathbf{q},\mathbf{0};\bm{\alpha})}{|\mathbf{q}|^4},
\end{align}
where the primed sum skips the singular term at $\mathbf{q}=\vzero$, and $A_\mathrm{d}(\mathbf{q},\mathbf{0};\bm{\alpha})$ is the chosen auxiliary function controlled by some parameters $\bm{\alpha}$ which are fitted to approximate $S_\mathrm{d}(\mathbf{q},\vzero,\vzero)$ as closely as possible. In particular, the optimal parameters $\bm{\alpha}_0$ are those that minimize the following loss function:
\begin{align}\label{eq:loss-fn-quartic-direct}
    \bm{\alpha}_0 = \argmin_{\bm{\alpha}}\primedsum_{\mathbf{q} \in \mathcal{K}_\vq + \mathbb{L}^* , |\mathbf{q}|<q_\textrm{cut}}\left[\frac{S_\mathrm{d}(\mathbf{q},\vzero,\vzero)-A_\mathrm{d}(\mathbf{q},\mathbf{0};\bm{\alpha})}{|\mathbf{q}|^4}\right]^2,
\end{align}
where $q_\textrm{cut}$ is the cutoff radius for the fitting, usually extending multiple Brillouin zones, but including far fewer than the full $\Nk\Ng$ $\vq$-points in a practical calculation. The choice of loss function is motivated by making the numerical quadrature as small as possible, leading to the smallest error prefactor. Later, after \cref{sec:shape-structure-factor}, we will assume $A_\mathrm{d}(\mathbf{q},\mathbf{0}; \bm{\alpha}) = -|\mathbf{q}|^4 h_{\mathrm{d},4}(\mathbf{q} ; \bm{\alpha})^2$, where $h_{\mathrm{d},4}(\mathbf{q} ; \bm{\alpha})$ is the decay function of $\mathbf{q}$ used for this term. This allows for the computation of the other corrections to be feasible.

\begin{figure}
  \centering
  \includegraphics[width=0.65\textwidth]{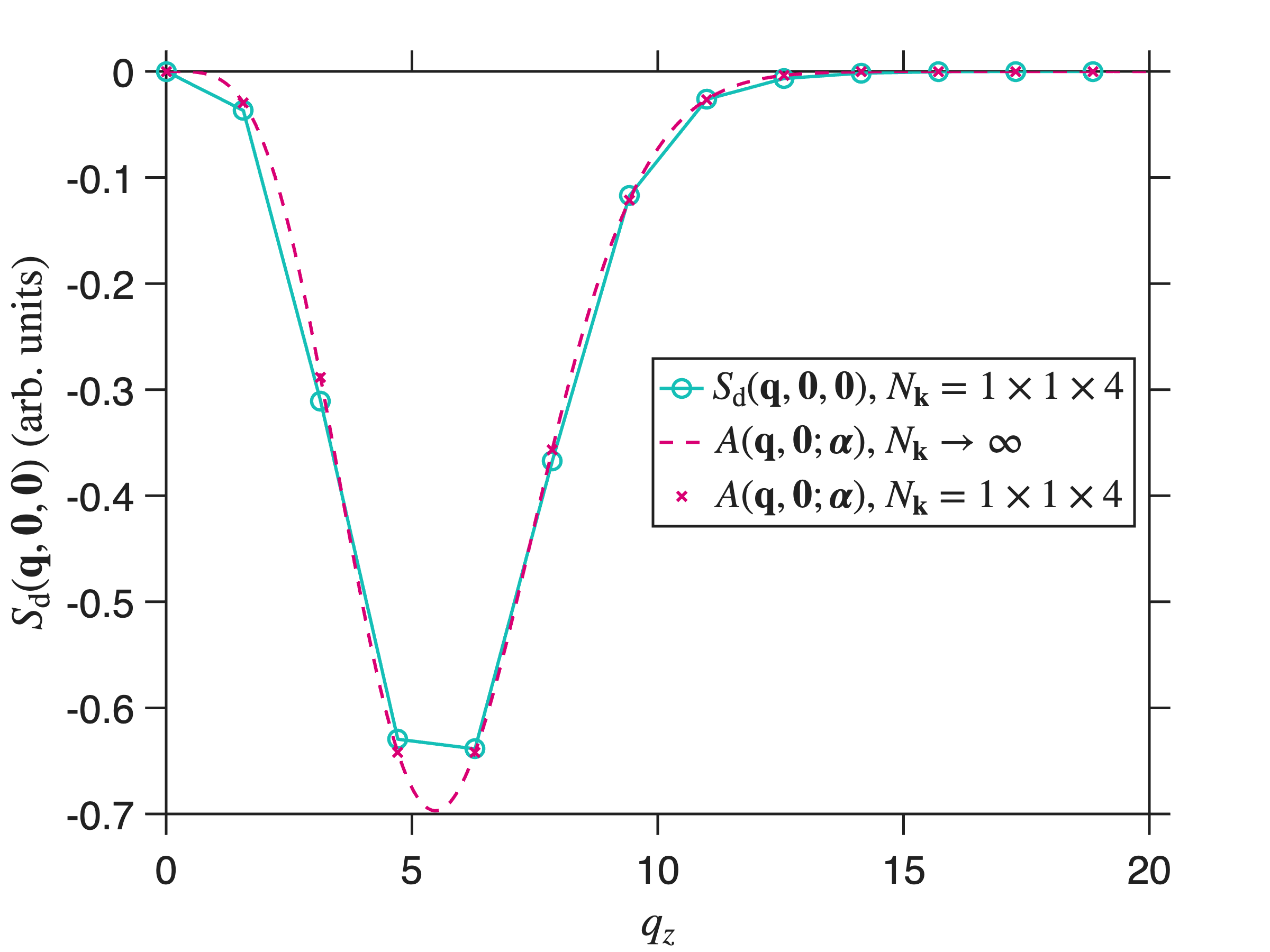}
    \caption{The quartic portion $S_{\mathrm{d}}(\mathbf{q},\vzero,\vzero)$ of the MP2 direct term structure factor for the model potential described in the text, plotted along a 1D path in reciprocal space. The fitted quartic auxiliary function $A(\mathbf{q},\vzero;\bm{\alpha})$ is also plotted}\label{fig:sqG_direct_1d_stacked_dG0}
\end{figure}

A visualization of the quartic portion of the structure factor $S_{\mathrm{d}}(\mathbf{q},\vzero,\vzero)$ and the corresponding fitted auxiliary function $A_\mathrm{d}(\mathbf{q},\mathbf{0};\bm{\alpha})$ is shown in Fig.\ref{fig:sqG_direct_1d_stacked_dG0} for a quasi-1D model system with the same isotropic Gaussian effective potential described in ref \cite{xingStaggeredMeshMethod2021}, except the well depth is raised from $-200$ to $-50$. In this example we use Gaussian decay, $h_{\mathrm{d},4}(\vq;\bm{\alpha})=\alpha_1 \exp\left(-\alpha_2|\vq|^2\right)$.
  
\paragraph{Second-order singularities in the direct term.}
These terms are slightly harder to correct, as care needs to be taken to avoid the naive $\mathcal{O}(\Nk^3\Ng^2)$ construction of $S_\mathrm{d}(\vq,\vzero,\Delta\vG)$ for all $\vq$ and $\Delta\vG$ suggested by the last term in \cref{eq:mp2_direct_tdl_deltaG_unfolded}. It is possible to motivate a form for the auxiliary function needed here, as we do later in \cref{sec:shape-structure-factor}, which suggests $A_\mathrm{d}(\mathbf{q},\Delta\mathbf{G};\bm{\alpha}') =
 |\mathbf{q}|^2 |\mathbf{q}+\Delta\mathbf{G}|^2 %
 h_{\mathrm{d},2}(\mathbf{q};\bm{\alpha}')h_{\mathrm{d},2}(\mathbf{q}+\Delta\mathbf{G};\bm{\alpha}')$. Here, the decay function $h_{\mathrm{d},2}(\mathbf{q};\bm{\alpha}')$ must decay to zero fairly quickly. The function depends on parameters $\bm{\alpha}'$ which are different from those used for $h_{\mathrm{d},4}(\mathbf{q};\bm{\alpha})$.

 Taking this idea and applying it along with singularity subtraction yields the following contribution to the MP2 energy
 \begin{align}\label{eq:quadratic-direct-contribution}
    E^{\textrm{SS}}_\textrm{MP2} \leftarrow \frac{1}{\Nk}&\primedsum_{\mathbf{q} \in \mathcal{K}_\vq + \mathbb{L}^*}\left(\frac{\tilde{S}_\mathrm{d,2}(\mathbf{q})}{|\mathbf{q}|^2}  - \frac{f_{\mathrm{d},2}(\mathbf{q}; \bm{\alpha}')}{|\mathbf{q}|^2}\right)
    + \frac{1}{|\Omega^*|} \sum_{\Delta\mathbf{G}\in\mathbb{L}^*\setminus \{\vzero\}} \int_{\mathbb{R}^3} d\mathbf{q} 
    h_{\mathrm{d},2}(\mathbf{q};\bm{\alpha}')h_{\mathrm{d},2}(\mathbf{q}+\Delta\mathbf{G};\bm{\alpha}'),
\end{align}
where $\tilde{S}_\mathrm{d,2}(\mathbf{q})$ is the ``folded'' version of the structure factor $S_\mathrm{d}(\mathbf{q},\vzero,\Delta\mathbf{G})$ defined as
\begin{align}
    \tilde{S}_\mathrm{d,2}(\mathbf{q}) &:= \sum_{\Delta\mathbf{G}\in\mathbb{L}^*\setminus \{\vzero\}} \frac{S_\mathrm{d}(\mathbf{q},\vzero,\Delta\mathbf{G})}{|\mathbf{q}+\Delta\mathbf{G}|^2}
\end{align}
and $ f_{\mathrm{d},2}(\mathbf{q};\bm{\alpha}')$ is the corresponding ``folded'' auxiliary function, defined in terms of $A_\mathrm{d}(\mathbf{q},\Delta\mathbf{G};\bm{\alpha}')$ as
\begin{align}
    f_{\mathrm{d},2}(\mathbf{q};\bm{\alpha}') &:=
    \sum_{\Delta\mathbf{G}\in\mathbb{L}^*\setminus \{\vzero\}}
    \frac{A_\mathrm{d}(\mathbf{q},\Delta\mathbf{G};\bm{\alpha}')}{|\mathbf{q}+\Delta\vG|^2}\\
    &=
    |\mathbf{q}|^2h_{\mathrm{d},2}(\mathbf{q};\bm{\alpha}')
    \left[\left(\sum_{\Delta\mathbf{G}\in\mathbb{L}^*} 
     h_{\mathrm{d},2}(\mathbf{q}+\Delta\mathbf{G};\bm{\alpha}') \right)-
     h_{\mathrm{d},2}(\mathbf{q};\bm{\alpha}') \right].\label{eq:fd2_expanded}
\end{align}
In these folded quantities, the sum excludes only the $\Delta\mathbf{G}=\vzero$ term, which is treated separately by the quartic correction above. Note, firstly, that the singularities at $\mathbf{q}=-\Delta\mathbf{G}$ are removed by the second-order vanishing of $S_\mathrm{d}(\mathbf{q},\vzero,\Delta\mathbf{G})$ and, secondly, that the quantity in the inner parentheses in \cref{eq:fd2_expanded} is periodic with respect to $\vq$ in $\Omega^*$. This folding of the structure factor and auxiliary function is an important step that allows efficient calculation of this correction, as it allows us to perform SS and fitting for all $\Delta\mathbf{G}$ terms at once, instead of separately for each $\Delta\mathbf{G}$, a procedure that would have ultimately scaled $\mathcal{O}(\Nk^3\Ng^2)$. Furthermore, we do not need to compute $\tilde{S}_\mathrm{d,2}(\mathbf{q})$ for all possible $\vq$, as the very first term in \cref{eq:quadratic-direct-contribution} is already included in standard MP2. We only need to compute $\tilde{S}_\mathrm{d,2}(\mathbf{q})$ for as many $\vq$ as are needed to obtain a good fit.

\begin{figure}
  \centering
  \includegraphics[width=0.65\textwidth]{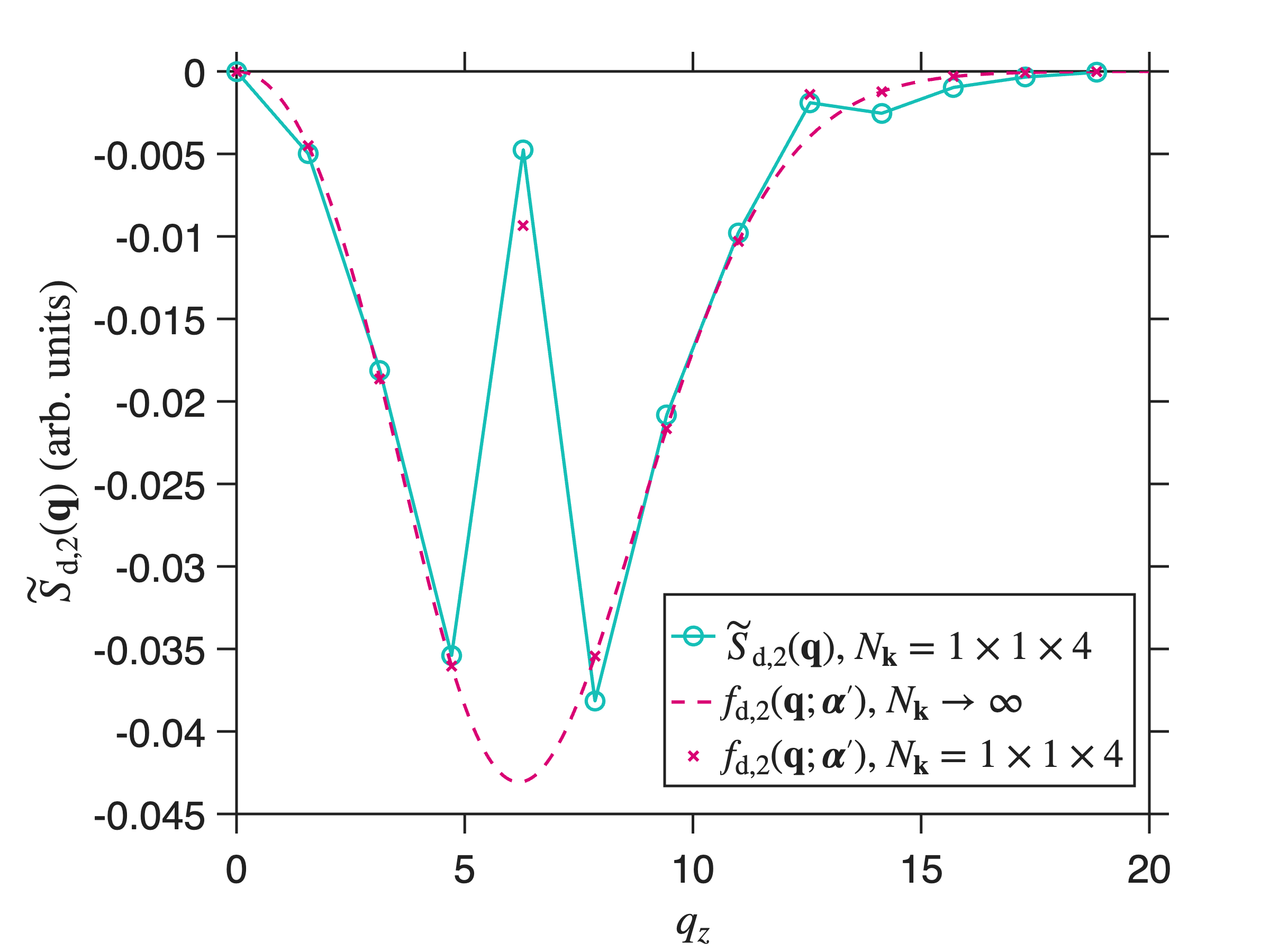}
  \caption{The quadratic portion $\tilde{S}_{\mathrm{d,2}}(\mathbf{q})$ of the MP2 direct term structure factor for the model potential described in the text, plotted along a 1D path in reciprocal space. The fitted quadratic auxiliary function $f_{\mathrm{d},2}(\mathbf{q};\bm{\alpha}')$ is also plotted, which attempts to capture the behavior of the structure factor including the discontinuities.}\label{fig:sqG_direct_1d_stacked_q2}
\end{figure}

Just as in the previous term, the optimal parameters $\bm{\alpha}'_0$ are those that minimize the following loss function:
\begin{align} \label{eq:loss-fn-quadratic-direct}
    \bm{\alpha}'_0 = \argmin_{\bm{\alpha}'} \primedsum_{\mathbf{q} \in \mathcal{K}_\vq + \mathbb{L}^*,|\vq|<q_\mathrm{cut}}\left[ \frac{\tilde{S}_\mathrm{d,2}(\mathbf{q})}{|\mathbf{q}|^2}  - \frac{f_{\mathrm{d},2}(\mathbf{q}; \bm{\alpha}')}{|\mathbf{q}|^2} \right]^2
\end{align}
 where the first term can be computed efficiently for essentially the same reason as for the second term above. The choice of this loss function is a careful tradeoff between minimizing the quadrature term and being computationally feasible (each $\Delta\mathbf{G}$ could be corrected separately, with different parameters, but the scaling of the method would be an unforgiving $\mathcal{O}(\Ng^2)$). Finally, an example of the folded structure factor $\tilde{S}_\mathrm{d,2}(\mathbf{q})$ and the corresponding folded auxiliary function $f_{\mathrm{d},2}(\mathbf{q};\bm{\alpha}')$ is shown in \cref{fig:sqG_direct_1d_stacked_q2} for the same quasi-1D model system as in the previous section. In this example we also use Gaussian decay, $h_{\mathrm{d},2}(\vq;\bm{\alpha}')=\alpha'_1 \exp\left(-\alpha'_2|\vq|^2\right)$.

\paragraph{Exchange Term.}
Just as in the previous term, it can be motivated (\cref{sec:shape-structure-factor}) that a good auxiliary function will look like $A_\mathrm{x}(\mathbf{q}_1,\mathbf{q}_2 ; \bm{\alpha}'')=|\mathbf{q}_1|^2 |\mathbf{q}_2|^2 h_{\mathrm{x}}(\mathbf{q}_1; \bm{\alpha}'') h_{\mathrm{x}}(\mathbf{q}_2; \bm{\alpha}'') \sim S_\mathrm{x}(\mathbf{q}_1,\mathbf{q}_2,\vzero,\vzero)$, where $h_{\mathrm{x}}(\mathbf{q}_1; \bm{\alpha}'')$ is the now-familiar general-form function fitted with parameters $\bm{\alpha}''$ different from those in the previous two subsections.

\begin{figure}
  \centering
  \includegraphics[width=0.65\textwidth]{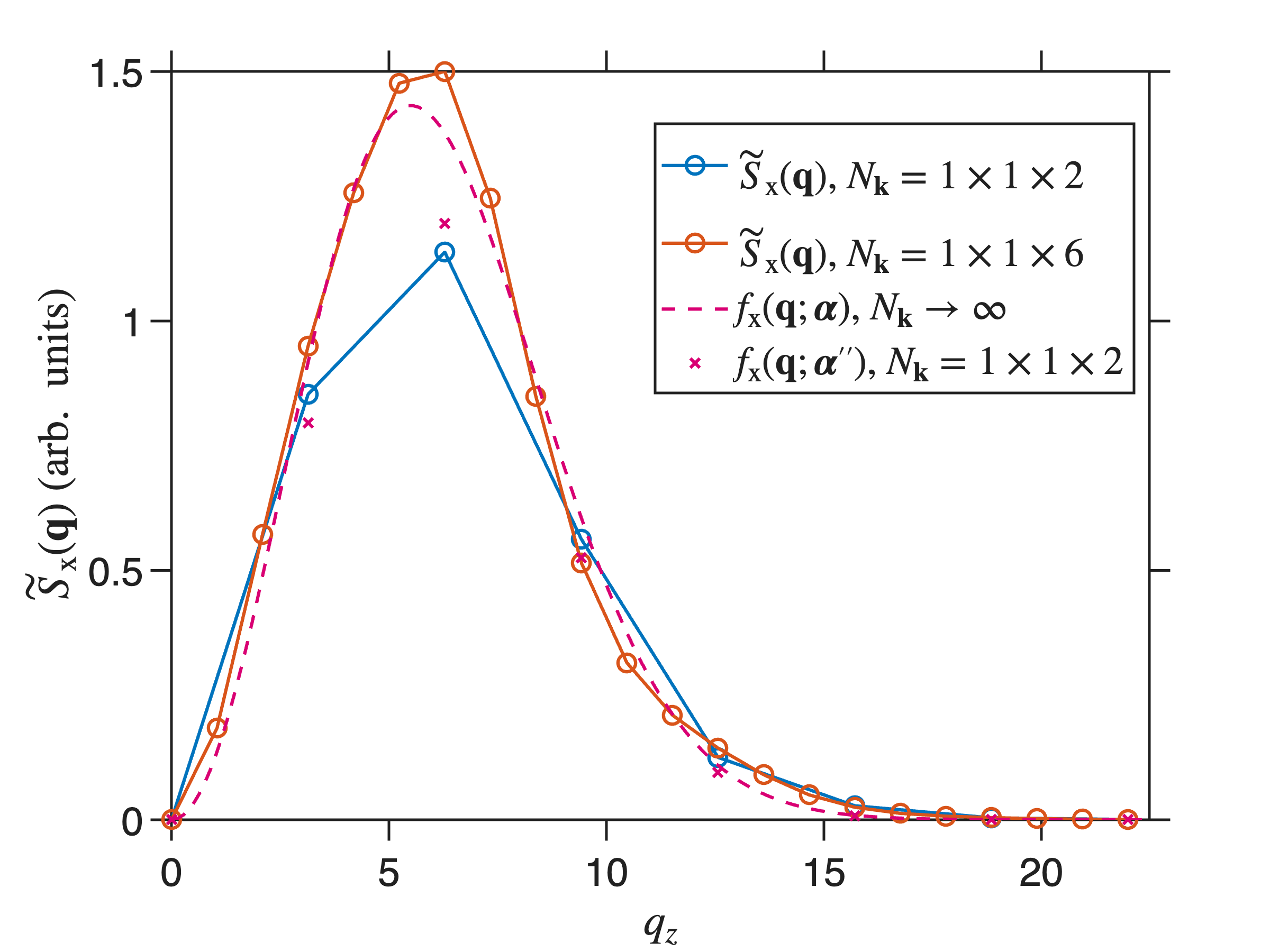}
  \caption{The folded exchange structure factor $\tilde{S}_{\mathrm{x}}(\mathbf{q}_1)$ for the model potential described in the text, plotted along a 1D path in reciprocal space. The fitted folded auxiliary function $f_{\mathrm{x}}(\mathbf{q}_1)$ is also plotted, which attempts to capture the behavior of the structure factor including the discontinuities.}\label{fig:sqG_exchange_1d_stacked}
\end{figure}

Using this approximate form for $A_\mathrm{x}(\mathbf{q}_1,\mathbf{q}_2 ; \bm{\alpha}'')$ and the idea of singularity subtraction, the following final contribution to the MP2 energy can be derived as 
\begin{align}\label{eq:exchange-contribution}
    E^{\textrm{SS}}_\textrm{MP2} \leftarrow \frac{1}{\Nk} & \left( 
    \primedsum_{\mathbf{q} \in \mathcal{K}_\vq + \mathbb{L}^*} \frac{\tilde{S}_\mathrm{x}(\mathbf{q})}{|\mathbf{q}|^2} -
    \primedsum_{\mathbf{q} \in \mathcal{K}_\vq + \mathbb{L}^*} \frac{f_\mathrm{x}(\mathbf{q};\bm{\alpha}'')}{|\mathbf{q}|^2}
    \right)
    +\frac{1}{|\Omega^*|^2}  \left(\int_{\mathbb{R}^3} d\mathbf{q} 
    h_{\mathrm{x}}(\mathbf{q};\bm{\alpha}'') \right)^2,
\end{align}
where the first term in the equation is already included in the standard MP2 calculation. Similar to the previous term, the structure factor $\tilde{S}_\mathrm{x}(\mathbf{q})$ is the folded version of $S_\mathrm{x}(\mathbf{q}_1,\mathbf{q}_2,\vzero,\vzero)$ defined as
\begin{align}
  \tilde{S}_{\mathrm{x}}(\mathbf{q}_1)
  :=
  \frac{1}{\Nk}\primedsum_{\mathbf{q}_2 \in \mathcal{K}_{\mathbf{q}} + \mathbb{L}^*}
  \frac{S_{\mathrm{x}}(\mathbf{q}_1,\mathbf{q}_2,\vzero,\vzero)}{|\mathbf{q}_2|^2}.
\end{align}
Its corresponding ``folded'' auxiliary function is defined as
\begin{align}
  f_{\mathrm{x}}(\mathbf{q}_1;\bm{\alpha}'')&:=\frac{1}{\Nk}
    \sum_{\mathbf{q}_2 \in (\mathcal{K}_{\mathbf{q}} + \mathbb{L}^*)\setminus\{\vzero\}}
    \frac{A_\mathrm{x}(\mathbf{q}_1,\mathbf{q}_2 ; \bm{\alpha}'')}{|\mathbf{q}_2|^2}\\
  &=\frac{|\mathbf{q}_1|^2
  h_{\mathrm{x}}(\mathbf{q}_1;\bm{\alpha}'')}{\Nk}
  \sum_{\mathbf{q}_2 \in (\mathcal{K}_{\mathbf{q}} + \mathbb{L}^*)\setminus\{\vzero\}}
  h_{\mathrm{x}}(\mathbf{q}_2;\bm{\alpha}'').
\end{align}
Once again, this folding allows us to efficiently perform SS for this term, as naively constructing and correcting 
$S_\mathrm{x}(\mathbf{q}_1,\mathbf{q}_2,\vzero,\vzero)$ would prohibitively scale $\mathcal{O}(\Nk^3\Ng^2)$. The optimal parameters $\bm{\alpha}''_0$ are those that minimize the following loss function:
\begin{align} \label{eq:loss-fn-exchange}
  \bm{\alpha}''_0 = \argmin_{\bm{\alpha}''} \primedsum_{\mathbf{q} \in \mathcal{K}_\vq + \mathbb{L}^*, |\mathbf{q}| < q_{\mathrm{cut}}} \left[ \frac{\tilde{S}_\mathrm{x}(\mathbf{q})}{|\mathbf{q}|^2} -
     \frac{f_{\mathrm{x}}(\mathbf{q};\bm{\alpha}'')}{|\mathbf{q}|^2} \right]^2
\end{align}
An example of the folded exchange structure factor and the fitted folded auxiliary function is shown in  \cref{fig:sqG_exchange_1d_stacked} for the same quasi-1D model system as in \cref{fig:sqG_direct_1d_stacked_dG0,fig:sqG_direct_1d_stacked_q2} and with the same Gaussian decay. The $\mathcal{O}(\Nk^{-1})$ errors for all $\vq$ in $\tilde{S}_\mathrm{x}(\mathbf{q})$ are clearly visible when comparing the structure factors at $\Nk=1\times1\times2$ and $\Nk=1\times1\times6$. The fitted auxiliary function is able to ``reproduce'' these discontinuities, which allows for a better approximation to the structure factor at the TDL and thus better FSE corrections.

\subsection{Choice of Auxiliary Functions} \label{sec:shape-structure-factor}
We have outlined in previous sections how knowledge of the criteria for successful uniform quadrature can be used within a simple correction scheme to produce theoretically sound and computationally feasible corrections. But these corrections rely on finding auxiliary functions $A_\mathrm{d}(\mathbf{q},\Delta\mathbf{G}; \bm{\alpha}), A_\mathrm{x}(\mathbf{q}_1,\mathbf{q}_2;\bm{\alpha})$ that resemble the structure factors $S_\mathrm{d}(\mathbf{q},\vzero,\Delta\mathbf{G}), S_\mathrm{x}(\mathbf{q}_1,\mathbf{q}_2,\vzero,\vzero) $. In this section we motivate the qualitative shape of the auxiliary functions by considering the symmetries and small $\mathbf{q}$ expansions of the structure factors. The simple form of the auxiliary functions obtained here also proves crucial in simplifying the expressions and developing a feasible MP2SS scheme (as seen in \cref{sec:MP2SS}).

The structure factors, as defined in \cref{eq:unfolded-direct-structure-factor-full-defition,eq:unfolded-exchange-structure-factor-full-definition}, have analytic asymptotic expansions. Consider a Taylor expansion of the reciprocal pair densities $\hat{\rho}_{i\mathbf{k}_i j (\mathbf{k}_i +\mathbf{q})}(\mathbf{G})$ defined in \cref{eq:pair_densities}. When the unfolded wavevector $\mathbf{q}+\mathbf{G}$ approaches $\vzero$, the zeroth-order term vanishes by orbital orthogonality, so the pair density vanishes linearly in $\mathbf{q}+\mathbf{G}$. Consequently, near the singular limits, $S_\mathrm{d}(\mathbf{q},\mathbf{G},\mathbf{G}')$ is second order in the small arguments $\mathbf{q}+\mathbf{G}$ and $\mathbf{q}+\mathbf{G}'$, and, similarly, $S_\mathrm{x}(\mathbf{q}_1, \mathbf{q}_2, \mathbf{G}_1, \mathbf{G}_2)$ is second order in $\mathbf{q}_1+\mathbf{G}_1$ and $\mathbf{q}_2+\mathbf{G}_2$.
Taking this small $\mathbf{q}$ behavior and considering the symmetries of the structure factor (\cref{subsec:singularities_classification}), it is possible to arrive at multiple expressions that satisfy both symmetry and asymptotics. But perhaps the simplest candidate expressions for the auxiliary functions are 
\begin{align}
    A_\mathrm{d}(\mathbf{q},\Delta\mathbf{G})  &= |\mathbf{q}|^2 |\mathbf{q}+\Delta \mathbf{G}|^2 h_{\mathrm{d}}(\mathbf{q})h_{\mathrm{d}}(\mathbf{q}+\Delta \mathbf{G}) \\
    A_\mathrm{x}(\mathbf{q}_1,\mathbf{q}_2 ) &=|\mathbf{q}_1|^2 |\mathbf{q}_2|^2 h_{\mathrm{x}}(\mathbf{q}_1) h_{\mathrm{x}}(\mathbf{q}_2)
\end{align}
where decay functions $h_{\mathrm{d}}, h_{\mathrm{x}}$ have been introduced. The prefactors to these decay functions capture the correct asymptotic behavior, while the decay functions ensure a smooth decay of the auxiliary functions at large $\mathbf{q}$. 

The theoretical guarantees of SS allow for imperfect auxiliary functions to still improve the rate of error convergence. But a good auxiliary function that accurately captures the structure factors will lead to an additional advantage: a small error prefactor on top of the improved convergence rate. For this reason, we give more flexibility to our auxiliary functions by furnishing the decay functions with optimizable parameters $\bm{\alpha}$, represented explicitly as $h_{\mathrm{d}}(\mathbf{q};\bm{\alpha})$, $h_{\mathrm{x}}(\mathbf{q};\bm{\alpha})$, as done previously  \cite{quitonOptimizedAuxiliaryFunctions2025}. 

Due to the complexity of real systems, which prevents acquiring concise closed-form expressions for the structure factors, the choice of decay functions must be done mostly empirically. 
Here, we present three possible configurations of MP2SS, each with their specific choice of auxiliary function decay, as outlined in \cref{tab:mp2ss_configurations}. 
All auxiliary functions across all configurations are fitted with the appropriate Coulomb-weighting for each term, as seen in \cref{eq:loss-fn-quartic-direct,eq:loss-fn-quadratic-direct,eq:loss-fn-exchange}. 

The first two configurations, MP2SS-Gaussian and MP2SS-Exponential, are the simplest, and use the same decay function for all terms. Gaussian decay reflects both the smoothness of the direct and exchange structure factors at the thermodynamic limit and the assumption that, because Gaussian basis sets are used, the structure factors should decay like Gaussians at large $\vq$. Exponential decay, on the other hand, allows for better agreement in the behavior of the auxiliary function near the origin at the cost of losing smoothness. Previous work \cite{quitonOptimizedAuxiliaryFunctions2025} has also shown that exponential decay may be appropriate for larger Gaussian basis sets, which can better capture non-smooth features like cusps and therefore result in structure factors which decay slower than a Gaussian.

We also present a third configuration, MP2SS-Tuned. This configuration uses different decay functions for each term to examine the possibility of achieving the TDL at the sparsest $\vk$-mesh by tailoring an auxiliary function to each term. One can justify the choices as follows. As observed in other work in FSE corrections for post-SCF methods \cite{liaoCommunicationFiniteSize2016}, very fine $\vk$-meshes are needed to resolve the smooth part of the structure factor at small $\vq$ in real materials. In our work, we find the fourth-order direct term especially suffers from this. Since in our usual computational $\vk$-meshes the smoothness of this term is not resolved, and it appears to have a cusp, we choose the exponential decay function to fit it. The second-order direct term is better resolved, so its smooth part is visible and a smooth auxiliary function is best used. Since the flexibility of a Gaussian seems enough to capture the shape of this term, we decide not to introduce a more complex auxiliary function. Finally, it seems the exchange term benefited from a slightly more flexible auxiliary function, so the quartic exponential function was introduced, mirroring our previous work \cite{quitonOptimizedAuxiliaryFunctions2025}.

\begin{table}[h!]
\centering
\caption{Summary of the decay functions used for the auxiliary functions for each MP2 term. The functions $h_{\mathrm{d},4}(\mathbf{q};\bm{\alpha})$, $h_{\mathrm{d},2}(\mathbf{q};\bm{\alpha}')$, and $h_{\mathrm{x}}(\mathbf{q};\bm{\alpha}'')$ correspond to the fourth-order singularity in the direct term, the second-order singularity in the direct term, and the exchange term, respectively, and they are all functions of $\mathbf{q}$ only through its magnitude $q=|\vq|$. The total number of optimizable parameters is 2 or 4 depending on the function, and they are labeled  $\alpha_i$. Each column represents a different collection of function choices making up one of the three tested configurations: MP2SS-Gaussian, MP2SS-Exponential or MP2SS-Tuned. }\label{tab:mp2ss_configurations}

\begin{tabular}{cccc}
\toprule
Decay Function & Gaussian Config. & Exponential Config. & Tuned Config. \\
\midrule
$h_{\mathrm{d},4}(\mathbf{q};\bm{\alpha})$
& $\alpha_1\exp\!\left(-\dfrac{q^2}{2\alpha_2^2}\right)$ 
& $\alpha_1 \exp(-\alpha_2 q)$ 
& $\alpha_1 \exp(-\alpha_2 q)$ \\

$h_{\mathrm{d},2}(\mathbf{q};\bm{\alpha}')$
& $\alpha_1 \exp\!\left(-\dfrac{q^2}{2\alpha_2^2}\right)$ 
& $\alpha_1 \exp(-\alpha_2 q)$ 
& $\alpha_1 \exp\!\left(-\dfrac{q^2}{2\alpha_2^2}\right)$ \\

$h_{\mathrm{x}}(\mathbf{q};\bm{\alpha}'')$
& $\alpha_1 \exp\!\left(-\dfrac{q^2}{2\alpha_2^2}\right)$ 
& $\alpha_1 \exp(-\alpha_2 q)$ 
& $\alpha_1 \exp(-\alpha_2 (\sqrt{1+\alpha_3 q^2+\alpha_4 q^4}-1))$\\
\bottomrule
\end{tabular}

\end{table}

The 1-particle basis also has a significant impact on the structure factor. For example, if a basis set is unable to capture the correct physical phenomena in the system (as is the case for systems with dispersive interactions in basis sets without polarization), the structure factor will not be converged, and it will look qualitatively different from that at the complete basis set limit. In general, we observe that there is a qualitative change of shape in the structure factor as the basis set is increased. This means that a particular auxiliary function will perform differently in different basis sets. In this paper, our objective is to be working as close to the MP2 basis-set limit as is computationally feasible. For a further discussion on the effect of the basis set on the structure factor and the performance of MP2SS, see Section~S1 in the Supporting Information (SI). %

\subsection{Computational Procedure}\label{subsec:computational_procedure}
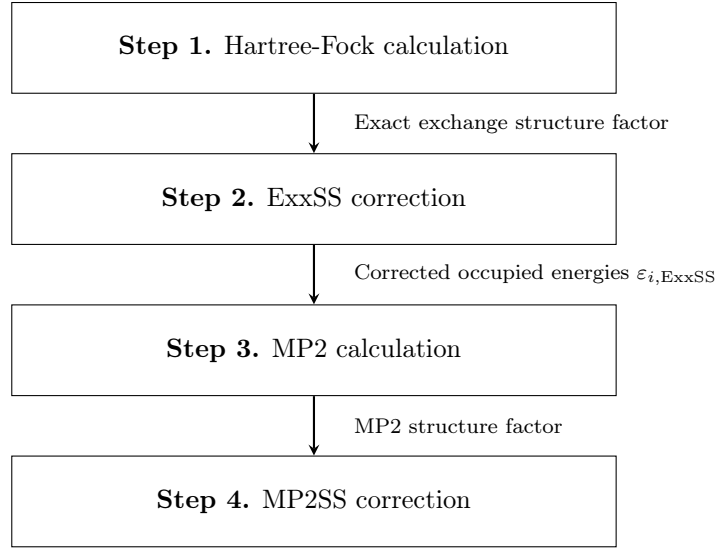
\begin{figure}[ht]
    \centering
    
    \begin{tikzpicture}[node distance=2.cm]
    
    \node (step1) [process] {\textbf{Step 1.} Hartree-Fock calculation};
    
    \node (step2) [process, below of=step1] {\textbf{Step 2.} ExxSS correction};
    
    \node (step3) [process, below of=step2] {\textbf{Step 3.} MP2 calculation};
    
    \node (step4) [process, below of=step3] {\textbf{Step 4.} MP2SS correction};
    
    \draw [arrow] (step1) -- node[midway, right=0.4cm] {\footnotesize Exact exchange structure factor} (step2);
    
    \draw [arrow] (step2) -- node[midway, right=0.4cm] {\footnotesize Corrected occupied energies $\varepsilon_{i,\mathrm{ExxSS}}$} (step3);
    
    \draw [arrow] (step3) -- node[midway, right=0.4cm] {\footnotesize MP2 structure factor} (step4);
    
    \end{tikzpicture}
    
    \caption{High-level flowchart to obtain the singularity subtracted MP2 correlation energy.}
    \label{fig:sqg_flowchart}
\end{figure}

After the choice of MP2SS configuration is made, the procedure to obtain the MP2SS correction is as follows: (1) perform a periodic HF calculation to obtain the orbitals and orbital energies; (2) compute the exact exchange structure factors for each term and fit the ExxSS auxiliary function to obtain the ExxSS correction to the occupied orbital energies (obtained by dividing the ExxSS correction by $\Nocc$); (3) perform a periodic MP2 calculation with the ExxSS-corrected orbital energies to obtain the MP2 structure factor; (4) fit the MP2SS auxiliary function to the MP2 structure factors via \cref{eq:loss-fn-quartic-direct,eq:loss-fn-quadratic-direct,eq:loss-fn-exchange} and evaluate the integral of the auxiliary function to obtain the MP2SS correction. A flowchart of this procedure is shown in Figure \ref{fig:sqg_flowchart}. For ExxSS, we follow the recommendations outlined in our previous work \cite{quitonOptimizedAuxiliaryFunctions2025} for the choice of auxiliary function (quartic-exponential) and fitting procedure. Numerical experiments on insulators (see Figures~S41 and~S42 in the SI) show that the corrections provided by ExxSS to the MP2 correlation energy are small compared to MP2SS, indicating that for such systems, ExxSS may not be necessary to achieve good convergence toward the TDL. ExxSS, however, is not the bottleneck in MP2SS. %

The walltime bottleneck of the MP2SS procedure is the building of the MP2 structure factors (for the purpose of fitting the auxiliary functions), scaling up to $\mathcal{O}(\Nk^{3}\Ng\Nocc^2 \Nvirt^2)$; specifically, the final contraction to obtain the structure factor at each $\vq+\vG$ point sampled scales naively as $\mathcal{O}(\Nk^{2}\Nocc^2 \Nvirt^2)$, and the number of $\vq+\vG$ points scales at worst $\mathcal{O}(\Nk\Ng)$. This cost can be lowered in several ways, first by reducing the number of sampled $\vq+\vG$ points. Unless noted otherwise, we only sample the points within a maximum $q$ radius, which is set to be twice that of the ExxSS structure factor fit radius, again following the recommendations from our previous work \cite{quitonOptimizedAuxiliaryFunctions2025}. Results are relatively insensitive to the choice of $q$ radius cutoff, and readers are directed to Figure~S39 in the SI for convergence tests. We can also consider sampling $\vq+\vG$ only along the reciprocal lattice vectors, which would reduce the runtime scaling to $\mathcal{O}(\Nk^{7/3}\Ng\Nocc^2 \Nvirt^2)$ assuming $\Nqg\sim N_\mathbf{k}$, instead of within a sphere of a certain cutoff radius. Line-sampling, while not providing as good a fit to the structure factor in principle, still allows us to fit to the diagonal terms of the Taylor expansion of the structure factors at $\vq+\vG=\vzero$. As shown in the results for diamond at the triple zeta level in Figure~S5, the FSE correction is not significantly affected compared to when sampling all $\vq+\vG$ within a ball. %

Because the pair densities in eq \ref{eq:pair_densities} are not constructed via Fast Fourier Transform (FFT), we can compute each of them with a real-space quadrature of our choice. The initial results in \cref{subsec:exploratory_calcs} were computed with a real-space FFT grid corresponding to a kinetic energy cutoff of 30 hartree, whereas the following results in \cref{subsec:results_extended_crystals,subsec:results_molecular_crystals} were computed with Becke level 0 grids \cite{takatsukaMinimaxApproximationDecomposition2008a,treutlerEfficientMolecularNumerical1995} provided in PySCF. Internal tests show that these grids are dense enough to compute the structure factor at a level sufficiently accurate for auxiliary function fitting, and that increasing the density does not improve performance.

Finally, for the triple-zeta set results \cref{subsec:results_extended_crystals,subsec:results_molecular_crystals}, we used two optimizations. First, we used the integral-direct range-separated density fitting (RSDF) implementation of periodic MP2 in ref \cite{bintrimIntegralDirectHartreeFock2022} for both the regular MP2 calculation and for MP2SS. Specifically, this allows for reduced memory scaling by recomputing the 2-electron-3-center (2e3c) density-fitting integrals, which are themselves blocked by the atomic orbitals (AOs) or MOs depending on the contraction and available disk space. See section 2 of \cite{bintrimIntegralDirectHartreeFock2022} for further details. Second, we can also exploit the Laplace-transform to factor out the energy denominator. This allows us to reduce the CPU-time scaling of the structure factor build from $\mathcal{O}(\Nk^{7/3}\Ng\Nocc^2 \Nvirt^2)$ to $\mathcal{O}(\Nk^{7/3}\Ng\Nocc^2 \Nvirt N_{\mathrm{Laplace}})$, where $N_{\mathrm{Laplace}}$ is the number of Laplace points, typically 10--15 for gapped systems. Readers are directed to Section~S2 in the SI for further details. %

\section{Results}\label{sec:results}
Unless otherwise stated, the calculation parameters are as follows. The unit cells for all extended solids, including diamond and silicon, are obtained from the Materials Project, which uses the PBE functional \cite{perdewGeneralizedGradientApproximation1996} for geometry optimization; the unit cells for all molecular crystals, including ammonia and urea are obtained from the X23 dataset \cite{reillyUnderstandingRoleVibrations2013,dolgonosRevisedValuesX232019}, which also uses the PBE functional with the TS dispersion correction \cite{tkatchenkoAccurateMolecularVan2009}. All calculations were performed with the PySCF package \cite{sunRecentDevelopmentsPySCF2020,sunPythonSimulationsChemistry2026}, using a kinetic energy cutoff of 56 hartree and the GTH-PBE pseudopotential and basis set \cite{goedeckerSeparableDualspaceGaussian1996}. All calculations here were performed with PySCF's default Gaussian Density Fitting (GDF) approach \cite{sunGaussianPlanewaveMixed2017,sunRecentDevelopmentsPySCF2020}. We use Monkhorst-Pack (MP) $\vk$-point grids that are $\Gamma$-centered for odd $\vk$-meshes and shifted by half a grid-space in each direction to sidestep the $\Gamma$-point for even $\vk$-meshes. As already mentioned, all test materials were limited to insulators and semiconductors, as MP2 diverges for metals.

\subsection{Exploratory Calculations}\label{subsec:exploratory_calcs}
We first show the effectiveness of the different MP2SS configurations for reducing the FSE in the canonical MP2 correlation energy for two solids and two molecular crystals using a minimal basis: diamond, silicon, ammonia, and urea. For these FSE curves, we define a reference TDL by fitting the two densest uncorrected data points to the leading-order form $E(N_{\vk}) = a\Nk^{-1}+b$ and taking $b$ as the extrapolated intercept. We indicate when the data is within millihartree accuracy per atom/monomer of this reference with a gray stripe. Because these sparse $\vk$-meshes might not yet be fully in the asymptotic regime, this extrapolated value should be interpreted as a practical benchmark, and log-log error plots are not presented. Regular FSE curves already highlight which methods perform well. Nonetheless, readers can consult Figure~S40 in the SI for an example log-log error plot. %

We also observed in these numerical experiments that the staggered mesh method appeared to approach a different limiting value than the other methods, especially for molecular crystals. This is most likely because the staggered mesh method implementation in PySCF at the time of writing computes the density matrix and orbital energies at both the shifted and unshifted $\vk$-points using a non-SCF band structure calculation that involves the construction of the Hartree operator $J$ and the exact-exchange operator $K$ via a truncated Coulomb potential. This truncated Coulomb kernel is only available with Fast Fourier Transform density fitting (FFTDF) in PySCF at the time of writing. Therefore, staggered mesh uses GDF for the two-electron integrals calculation and FFTDF for the density matrix, as opposed to all other results that use \textit{only} GDF. This causes staggered mesh to exhibit a slightly different apparent limit if the auxiliary basis is not large enough. Thus, for all results shown below (except for diamond with the GTH-SZV basis), the staggered mesh curve is shifted by a constant offset so that its densest available $\vk$-mesh value coincides with the extrapolated all-GDF reference. This alignment is used only to compare the decay of the FSE; it should not be interpreted as showing agreement in the absolute energies or in the true limiting value. For the unshifted curves, readers are directed to Section~S3 in the SI. Based on those unshifted curves, the difference between the apparent limiting values of the staggered-mesh and all-GDF data appears to be at most a few millihartree per atom/monomer. %

\begin{figure}[H]
    \centering
    \includegraphics[width=\textwidth]{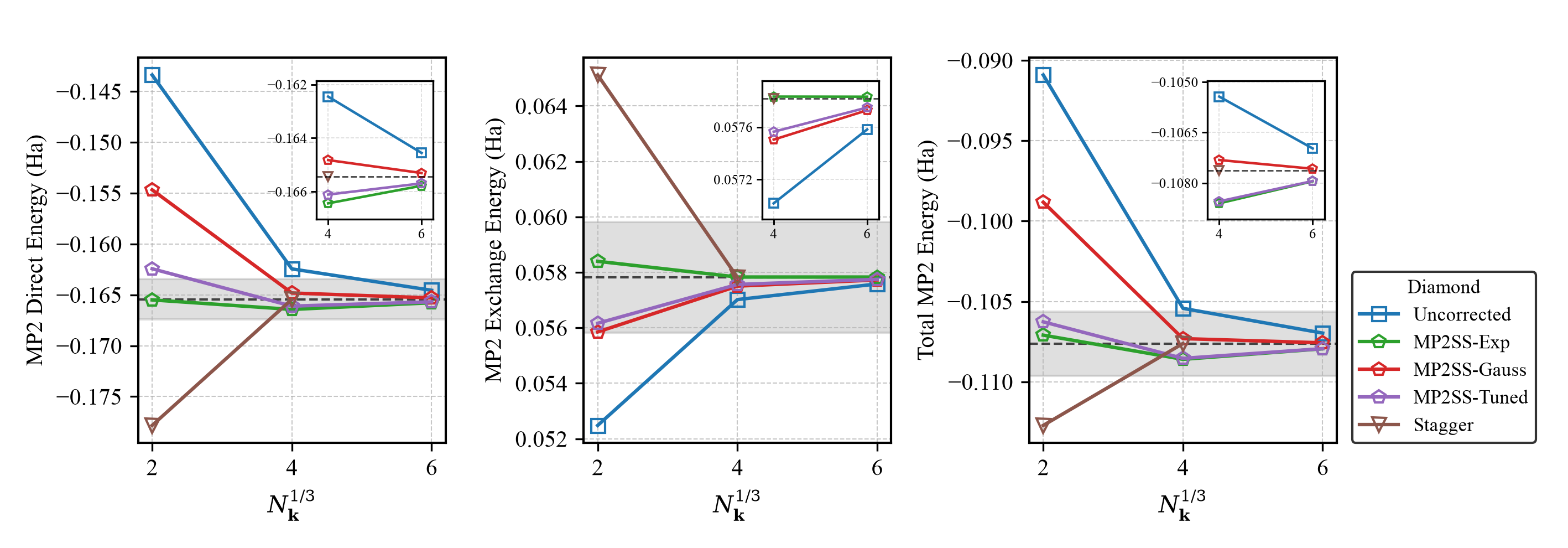}
    \caption{Finite-size effect of the direct, exchange, and total MP2 correlation energy per unit cell with the GTH-SZV basis set for diamond with different MP2SS configurations. The results for the (shifted) staggered mesh method are also included for comparison. The gray horizontal stripe represents millihartree accuracy per atom with respect to the fitted TDL reference obtained from an $\Nk^{-1}$ fit to the two densest points of the uncorrected curve. The $\Nk=6\times6\times6$ point for staggered mesh is missing due to disk space constraints from building the GDF Cholesky vectors.}
    \label{fig:diamond_szv_mp2ss_fse}
\end{figure}
\begin{figure}[H]
    \centering
    \includegraphics[width=\textwidth]{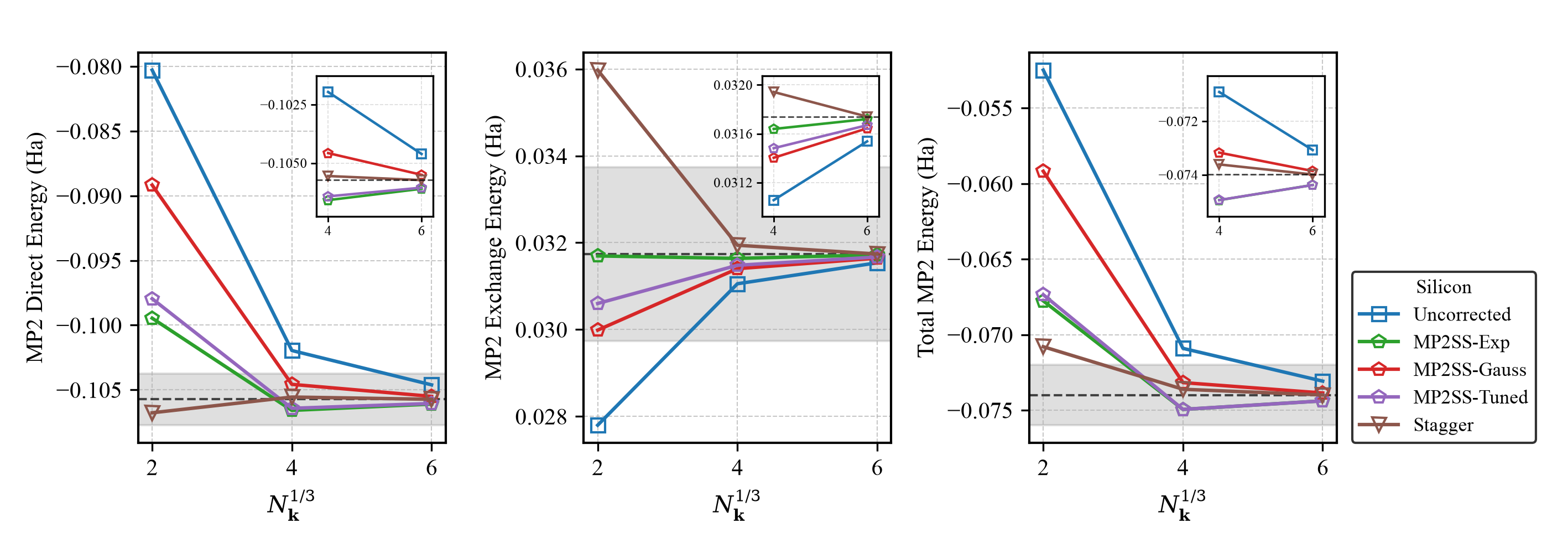}
    \caption{Finite-size effect of the direct, exchange, and total MP2 correlation energy per unit cell with the GTH-SZV basis set for silicon with different MP2SS configurations. The gray horizontal stripe represents millihartree accuracy per atom with respect to the fitted TDL reference obtained from an $\Nk^{-1}$ fit to the two densest points of the uncorrected curve.}
    \label{fig:silicon_szv_mp2ss_fse}
\end{figure}
\begin{figure}[H]
    \centering
    \includegraphics[width=\textwidth]{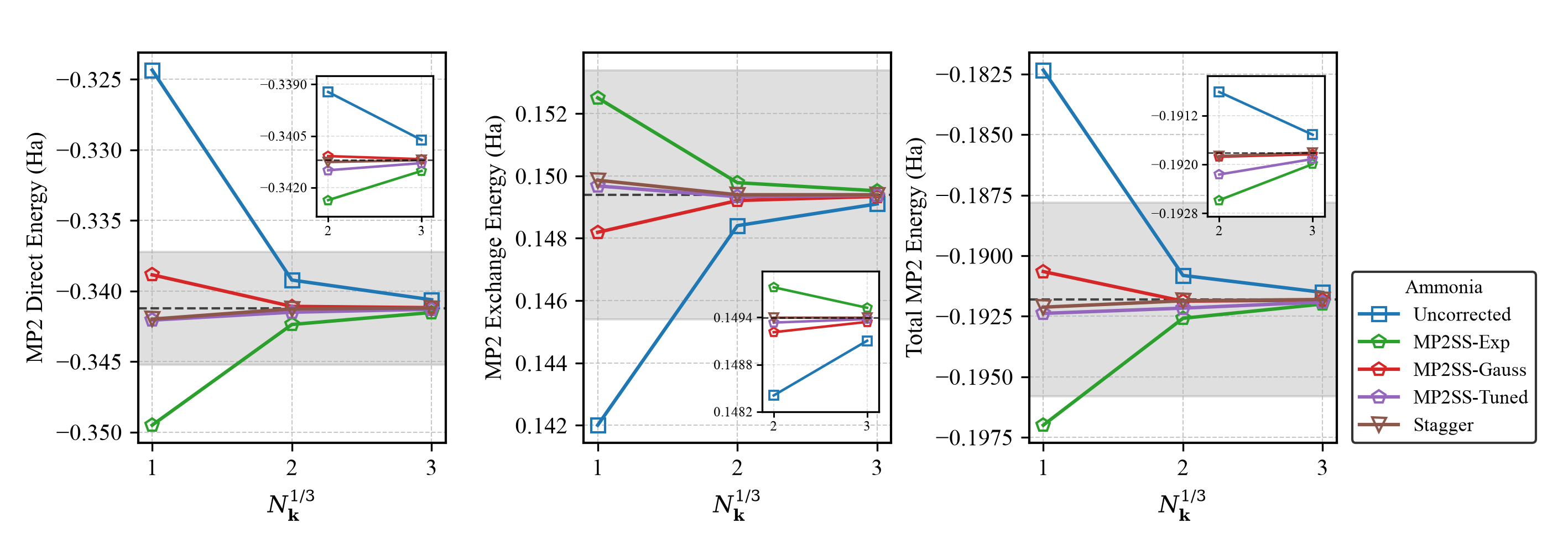}
    \caption{Finite-size effect of the direct, exchange, and total MP2 correlation energy per unit cell with the GTH-SZV basis set for ammonia with different MP2SS configurations. The gray horizontal stripe represents millihartree accuracy per monomer with respect to the fitted TDL reference obtained from an $\Nk^{-1}$ fit to the two densest points of the uncorrected curve.}
    \label{fig:ammonia_szv_mp2ss_fse}
\end{figure}
\begin{figure}[H]
    \centering
    \includegraphics[width=\textwidth]{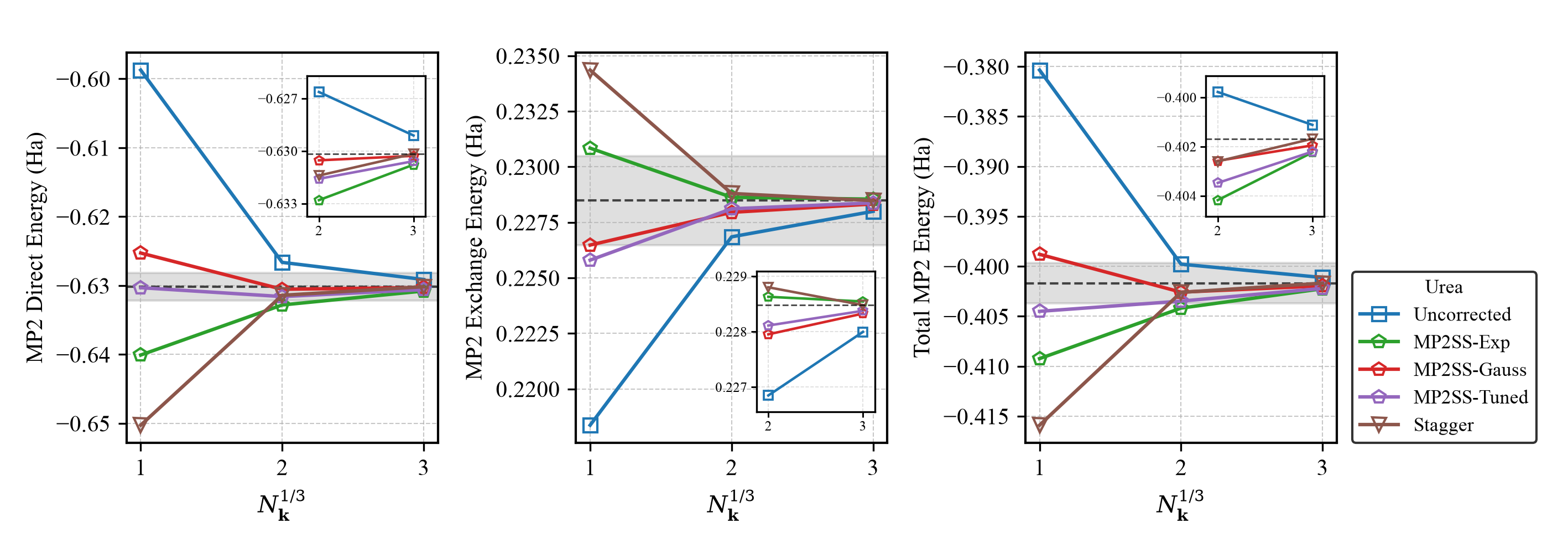}
    \caption{Finite-size effect of the direct, exchange, and total MP2 correlation energy per unit cell with the GTH-SZV basis set for urea with different MP2SS configurations. The gray horizontal stripe represents millihartree accuracy per monomer with respect to the fitted TDL reference obtained from an $\Nk^{-1}$ fit to the two densest points of the uncorrected curve.}
    \label{fig:urea_szv_mp2ss_fse}
\end{figure}

We first discuss the performance of the three MP2SS configurations for diamond, silicon, ammonia, and urea at the GTH-SZV basis set as shown in \cref{fig:diamond_szv_mp2ss_fse,fig:silicon_szv_mp2ss_fse,fig:ammonia_szv_mp2ss_fse,fig:urea_szv_mp2ss_fse}. The trend in terms of the corrections that MP2SS provides on top of uncorrected MP2 for each term appears to follow an order, with MP2SS-Gaussian providing the weakest correction, followed by MP2SS-Tuned, and then finally MP2SS-Exponential with the strongest. MP2SS-Exponential frequently over-corrects the standard MP2 correlation energy components, with the exception of diamond and silicon at the sparsest $\vk$-mesh values; conversely, MP2SS-Gaussian appears to provide an under-correction in all scenarios. MP2SS-Tuned, which contains a mixture of auxiliary functions from the prior two configurations, often (but not always) lands between MP2SS-Gaussian and MP2SS-Exponential.

The respective performances of MP2SS-Gaussian and MP2SS-Exponential across this test set may be rationalized by the role that the decay function plays in the small-$q$ behavior of the Coulomb-weighted quantities entering the loss functions. A Gaussian decay is an even, smooth function of $q$, whereas $\exp(-\alpha q)$ has a cusp at $q=0$ and contains both even and odd powers in its one-sided expansion. The exponential form therefore offers fitting flexibility near the origin, which could benefit unit cells where the $\vq$-grid density does not resolve the smoothness of the structure factor near the origin. However, on large unit cells whose structure factors are sufficiently smooth and even at the resolution of the sparsest $\vk$-meshes, the exponential decay function can also be more prone to overcorrection. In contrast to MP2SS-Gaussian and MP2SS-Exponential, the MP2SS-Tuned configuration uses decay functions and fitting procedures that are chosen for each MP2 term based on how well they correct that specific term at the sparsest $\vk$-mesh rather than \textit{a priori} estimations of how the auxiliary function should behave near the origin. As such, while MP2SS-Tuned does reach the TDL earlier than the other two configurations, the performance at larger $\vk$-meshes appears to be similar for all configurations. In fact, \cref{fig:ammonia_szv_mp2ss_fse,fig:urea_szv_mp2ss_fse} suggests that when the structure factor smoothness is sufficiently resolved, as is the case for larger unit cells, MP2SS-Gaussian performs the best. 

The results in this section begin to suggest that MP2SS-Gaussian provides the most robust FSE corrections for the correlation energy due to the smoothness of the auxiliary function near the origin. However, in order for MP2SS-Gaussian to perform well, the $\vk$-point mesh must be dense enough to ensure not only that the 1-particle density matrix is converged (as discussed in \cref{sec:shape-structure-factor}) but also that the smoothness of the Taylor expansion near the origin is sufficiently resolved. A $2\times2\times2$ $\vk$-point mesh for diamond and silicon, for example, is insufficient at the MP2 level (and even at the HF level). That same mesh, however, would suffice for urea and ammonia. We also note that the ability to resolve the structure factor behavior near the origin has also been previously observed and discussed in other periodic post-SCF calculations \cite{liaoCommunicationFiniteSize2016}.

\subsection{Extended Crystals}\label{subsec:results_extended_crystals}
We now present results with a triple-zeta quality basis set, for six extended solids, including those that are covalent (diamond, silicon), ionic (LiF, LiH, MgO), and anisotropic (hexagonal BN). These calculations were performed with the GTH-cc-pVTZ-lc basis, its corresponding auxiliary basis, and the GTH-HF-rev pseudopotential \cite{yeCorrelationConsistentGaussianBasis2022}. In order to make these calculations feasible, we sampled the structure factors along the reciprocal lattice vectors. Tests on diamond (see Section~S4 in the SI) show that this method of sampling has little effect on the subsequent MP2SS correction. We also use the Laplace-Transform and integral-direct to accelerate the structure-factor build as discussed in \cref{subsec:computational_procedure}. Finally, for hexagonal BN, we used $2\times2\times1$, $4\times4\times2$, and $6\times6\times3$ $\vk$-meshes to account for the shorter lattice dimension along the $z$-direction, and we used a $\Gamma$-centered mesh for all $\Nk$ since the last $\vk$-mesh has an odd number of $\vk$-points along $k_z$. %

\begin{figure}[!htbp]
    \centering
    \includegraphics[width=\textwidth]{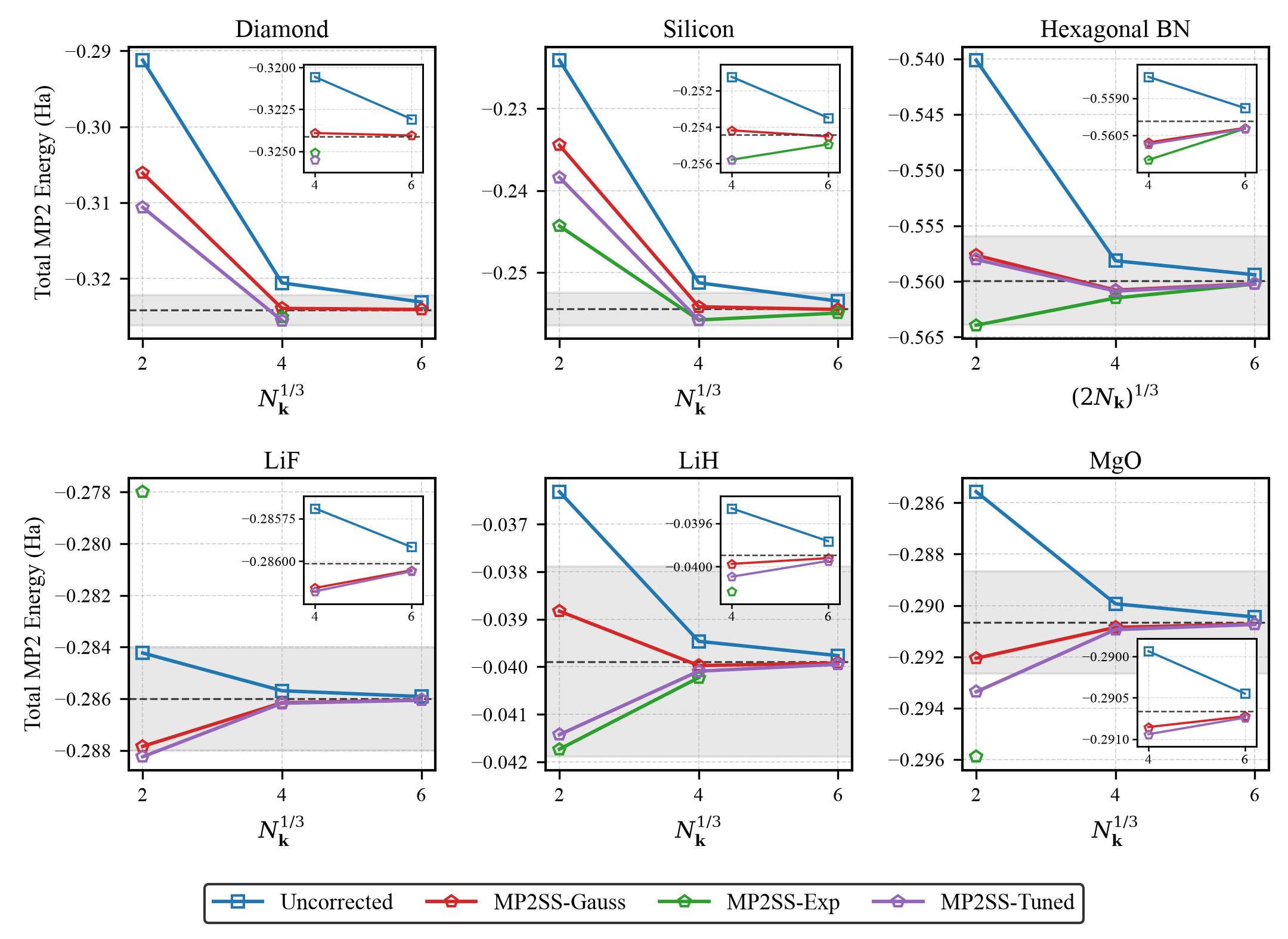}
    \caption{Total MP2 correlation energy for diamond, silicon, hexagonal BN, LiF, LiH, and MgO as a function of the linear \(\vk\)-mesh size. Uncorrected $\vk$-points MP2 (KMP2) results are compared with MP2SS-Gauss, MP2SS-Exp, and MP2SS-Tuned, and the dashed horizontal lines show the fitted TDL reference energies. The gray horizontal bands on the main panels represent millihartree accuracy per atom with respect to those references; the bands are omitted from the insets, which enlarge the results at the two densest available \(\vk\)-meshes. Plot points for some MP2SS-Exp lines are missing due to clear auxiliary function fitting failures. See the available data in the Supplementary Information.}
    \label{fig:simple_crystals_total_mp2_fse}
\end{figure}

The FSE corrections to the MP2 correlation energy, shown in  \cref{fig:simple_crystals_total_mp2_fse}, corroborate the findings in the previous section regarding the performance of the different MP2SS configurations. None of the configurations appear to get close enough to the estimated TDL at the smallest $\vk$-mesh, but at $4\times4\times4$ (or $4\times4\times2$ for hexagonal BN), the benefit of the correction becomes much clearer. MP2SS-Gaussian consistently performs the best in this regime. Several plot points for MP2SS-Exponential are missing due to clearly infeasible MP2SS corrections, pointing to robustness issues with fitting a non-smooth auxiliary function. See the data in the provided Supplementary Information for further details. MP2SS-Tuned, while still providing an improvement over uncorrected MP2, does not convincingly outperform MP2SS-Gaussian when the $\vk$-mesh is dense enough.

\subsection{Molecular Crystals}\label{subsec:results_molecular_crystals}
Finally, we present results with a triple-zeta quality basis set for all crystals in the X23 data set \cite{tkatchenkoAccurateMolecularVan2009}. Having observed its robustness for larger unit cells, we select MP2SS-Gaussian as the configuration for this benchmark. We use the same MP2 calculation parameters as ref \cite{liangCanSpinComponentScaled2023}. Specifically, we use the all-electron cc-pVTZ basis set  \cite{dunningGaussianBasisSets1989}, but use frozen core for the MP2 portion. Ref \cite{liangCanSpinComponentScaled2023} also provides estimates of the TDL for all 23 crystals, which we use as the reference for the results in this section. Here, we also use the Laplace-Transform and integral-direct routines to make the structure-factor builds feasible as discussed in \cref{subsec:computational_procedure}.

\begin{figure}[!htbp]
    \centering
    \includegraphics[width=\textwidth]{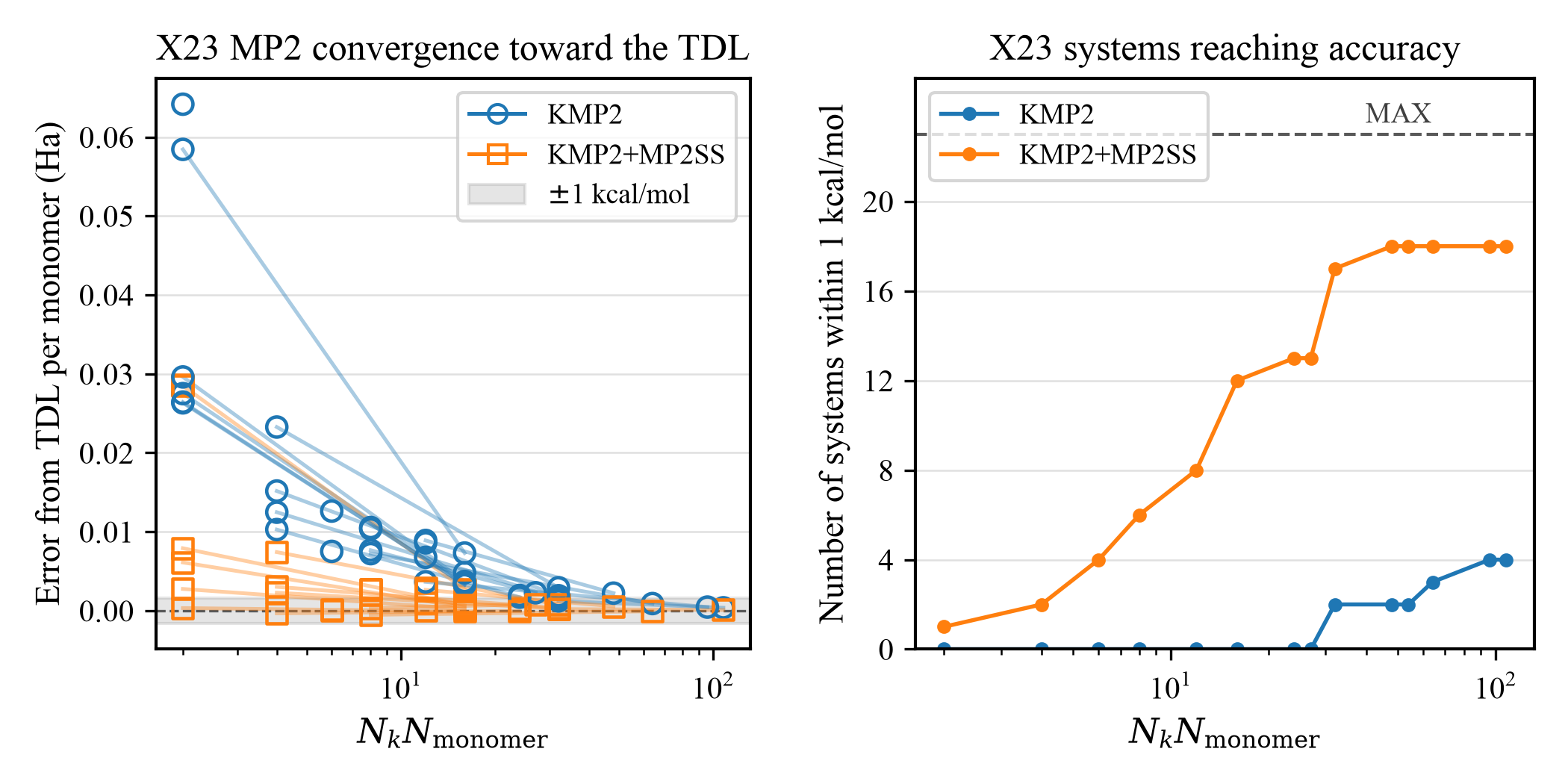}
    \caption{Convergence of uncorrected MP2 and MP2SS-corrected MP2 correlation energies across the X23 molecular-crystal benchmark as the number of sampled \(\vk\)-points increases. Left: error in the MP2 correlation energy per monomer relative to the TDL, estimated by ref \cite{liangCanSpinComponentScaled2023}; the gray band marks the 1 kcal/mol chemical-accuracy threshold. $N_{\mathrm{monomer}}$ is the number of monomers in the primitive unit cell. Each solid line connects the same crystal. Right: number of X23 systems within the 1 kcal/mol threshold.}
    \label{fig:x23_benchmark_convergence}
\end{figure}

The FSE plots for all 23 crystals are overlaid on top of each other on the left subplot of \cref{fig:x23_benchmark_convergence}. Comparing the blue and orange plot points, we see that MP2SS clearly reduces the error for all crystals and allows them to reach the TDL earlier than uncorrected MP2. The right subplot of \cref{fig:x23_benchmark_convergence} shows that as $\Nk$ is increased, the number of X23 crystals whose MP2 correlation energy per monomer is within chemical accuracy of the TDL is much higher when using MP2SS, with 18 of the crystals reaching chemical accuracy when $\Nk N_{\mathrm{monomer}}\approx100$. Uncorrected $\vk$-points MP2 (KMP2) only achieves this for 4 crystals. We note that the orange-line is flat toward the end of the curve because these calculations were given a seven-day time-limit, and therefore there are five systems---anthracene, imidazole, naphthalene, pyrazole, and uracil---whose regular MP2SS calculations did not finish at their respective larger $\vk$-meshes. If they finished and resulted in chemically accurate corrections, the orange line would be raised to 23 systems.

\begin{figure}[!htbp]
    \centering
    \includegraphics[width=\textwidth]{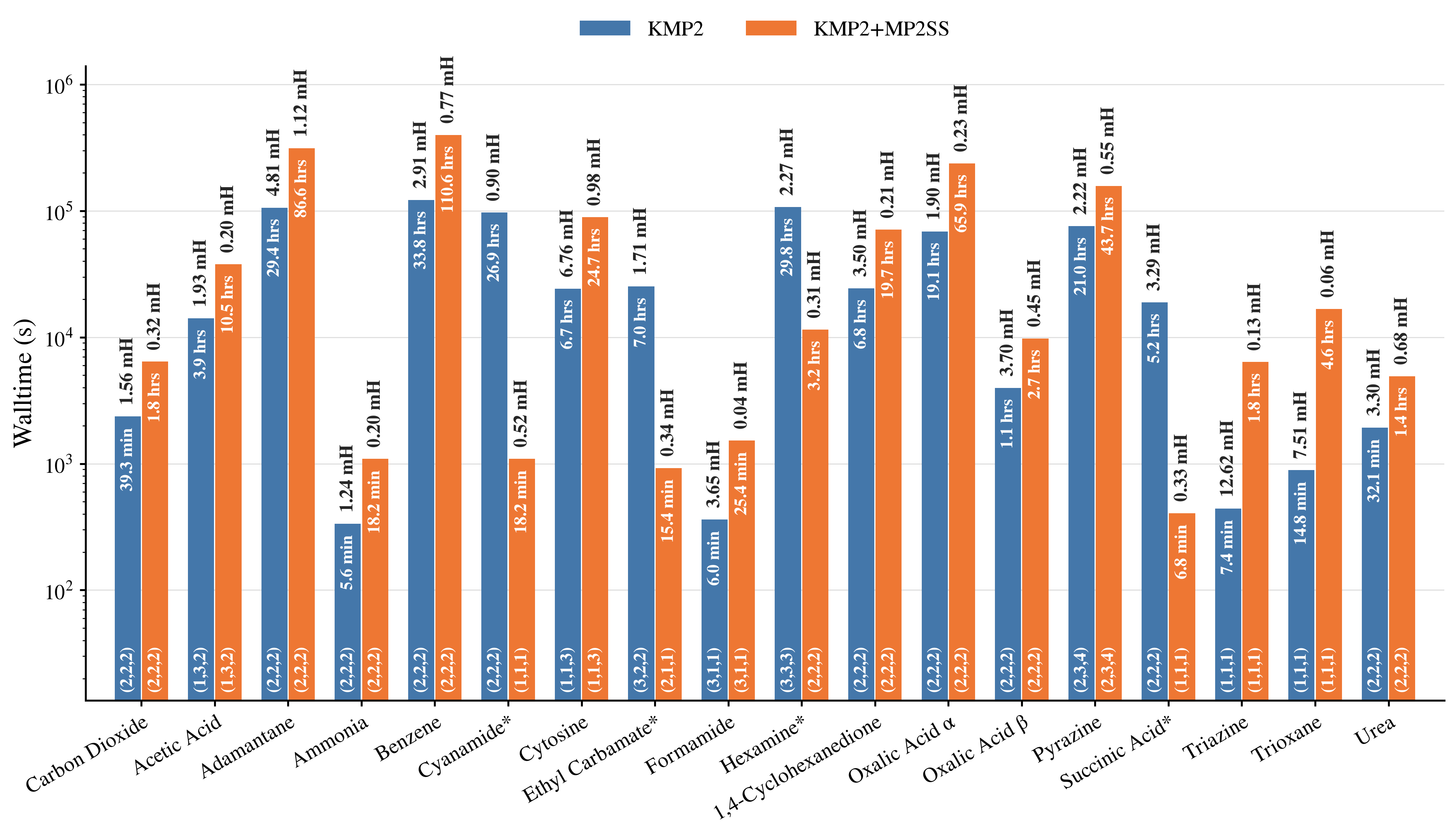}
    \caption{Wall time required by uncorrected KMP2 and MP2SS-corrected KMP2 calculations for the X23 systems where MP2SS reached chemical accuracy. The $\vk$-meshes were chosen based on the smallest mesh to reach kcal/mol accuracy ($\sim$1.594 millihartree); otherwise, the largest mesh that completed in 7 days is shown. The crystals where MP2SS achieved chemical accuracy at a smaller $\vk$-mesh is denoted by an asterisk (*). Each calculation was performed on a single node with an Intel Xeon Gold 6330, using 32 threads and 240 GB memory. Labels within the vertical bars give the \(\vk\)-point mesh and wall time; labels above the bars give the resulting error per monomer relative to the estimated TDL in millihartree. All calculations used the Laplace-transform and integral-direct routines discussed in \cref{subsec:computational_procedure} to allow for feasible structure factor calculations.}
    \label{fig:x23_walltime_accuracy}
\end{figure}

We also provide benchmark timings for the MP2SS routine in \cref{fig:x23_walltime_accuracy} for systems whose $\vk$-point meshes have reached chemical accuracy. For all systems shown, we see two cases: (1) a 5--15$\times$ reduction in FSEs for 3--5$\times$ wall time prefactor, or (2) in the case of cyanamide, ethyl carbamate, formamide, and hexamine, a 2--10$\times$ reduction in FSEs in addition to at least an order of magnitude reduction in wall time. We note that the KMP2-only run time is very fast for triazine and trioxane because they were performed at the $\Gamma$-point, and the RSDF integral-direct routine \cite{liangCanSpinComponentScaled2023} has optimizations for this case. These timing results underscore the effectiveness in MP2SS in obtaining TDL-converged MP2 correlation energies with, in the worst case, only a prefactor of a regular MP2 calculation wall time.

\section{Discussion}\label{sec:discussion}
This work demonstrates that the singularity subtraction approach---previously shown to be effective for exact exchange in periodic HF and hybrid DFT calculations---can be extended to MP2 calculations on real systems. By explicitly accounting for the singularities and discontinuities in the correlation energy structure factor through carefully chosen auxiliary functions, MP2SS reduces the prefactor of the finite-size error compared to uncorrected calculations and provides an alternative to the staggered mesh method without the need for additional density and orbital energy calculations at shifted $\vk$-meshes. The folded auxiliary function approach introduced in this work allows SS to directly address the discontinuities in the MP2 structure factor and obtain better fit parameters and provide accurate corrections even when individual structure factor components are not well-resolved at finite $\Nk$.

Among the three MP2SS configurations tested, MP2SS-Gaussian provides the most robust performance across our test set of extended solids and molecular crystals, often achieving millihartree and chemical accuracy at coarser $\vk$-point meshes. The MP2SS-Exponential consistently over-corrects the FSE except in cases where the Taylor expansion of the structure factor at the origin is not sufficiently resolved. While MP2SS-Tuned does appear to have good corrections at the sparsest $\vk$-meshes for some systems, it also begins to overcorrect like MP2SS-Exponential at larger $\vk$-meshes. We therefore recommend MP2SS-Gaussian as the default for routine calculations.

Further improvements can still be made to the MP2SS methodology. First, the ability to directly compute the first few derivatives of the structure factor at the origin could enhance the accuracy of the auxiliary-function fit for MP2SS-Gaussian, particularly for systems where this small-$\vq$ expansion is not resolved by the accessible $\vk$-meshes, as expected for low band-gap materials and small primitive unit-cells, such as silicon. Initial numerical tests suggest that the performance of MP2SS would improve if small-$\vq$ values of the structure factor could be reliably sampled. Second, as briefly mentioned in the results and computational procedure sections, we could also improve how the structure factor is built and sampled. The quadrature corrections and auxiliary-function integrals in eq \eqref{eq:singularity-subtraction-general} are negligible in all of the above calculations; the dominant additional cost comes instead from constructing the MP2 structure-factor data used in the fits.

The singularity subtraction framework developed here can be extended to other post-SCF theories, including MP3 and coupled-cluster theories, which exhibit similar singularities in their correlation energy expressions \cite{xingUnifiedAnalysisFinitesize2024,xingInverseVolumeScaling2024} but with additional challenges. For instance, finite-size errors for coupled-cluster come not only from the Coulomb kernel singularity but also from the self-consistent Newton iteration procedure, which propagates those errors into the coupled-cluster amplitudes themselves. This challenge must be addressed for singularity subtraction (or any finite-size error correction routine) to be effective. Singularity subtraction for a single-shot Green's function method (G0W0) is a possible future direction, as the screened Coulomb kernel also exhibits reciprocal-space singularities \cite{zhuAllElectronGaussianBasedG02021}.

 Furthermore, regularized MP2 methods, such as size-consistent Brillouin-Wigner perturbation theory (BW-s2)\cite{carter-fenkRepartitionedBrillouinWignerPerturbation2023,chenRegularizedPerturbationTheory2025}, also stand to benefit from this approach. Because MP2 diverges for metals and performs poorly for low band-gap materials, BW-s2 with the SS scheme in particular presents a promising way to accurately calculate the correlation energy for these challenging systems.

The development of MP2SS could also reduce the cost of generating full-system MP2-level training data with smaller finite-size errors for the fine-tuning of machine-learned interatomic potentials (MLIPs). In comparison, recent studies of MLIPs trained to post-SCF data frequently use local correlation, cluster or fragment-based approaches (not periodic boundary conditions), or are only able to sample sparse $\vk$-meshes due to computational cost \cite{meszarosShortRangeDMachineLearning2025,ikedaMachineLearningInteratomicPotentials2026,chenDataEfficientMachineLearning2023,schonbauerMachineLearnedForceFields2026,oneillRoutineCondensedPhase2025}. FSE corrections, like MP2SS, are potentially crucial in allowing these MLIPs to provide accurate ground state energies for large unit cells.

\section*{Acknowledgements}

This material is based upon work supported by the National Science Foundation Graduate Research Fellowship Program under Grant No. DGE 2146752 (SJQ) and the U.S. Department of Energy, Office of Science, Office of Advanced Scientific Computing Research and Office of Basic Energy Sciences, Scientific Discovery through the Advanced Computing (SciDAC) program under Award Number DE-SC0022364 (LL, MHG). LL is a Simons Investigator in Mathematics. This research also used the Savio computational cluster resource provided by the Berkeley Research Computing program at the University of California, Berkeley (supported by the UC Berkeley Chancellor, Vice Chancellor for Research, and Chief Information Officer).

\section*{Supporting information}
The Supporting Information is available free of charge via the Internet at \url{http://pubs.acs.org}. The Supporting Information PDF contains a discussion of the basis-set dependence of the MP2 structure factor; the minimax Laplace-transform formulation and contractions used to accelerate its construction; FSE curves with the staggered-mesh results shown without a vertical shift; validation of line sampling; separate direct, exchange, and total FSE curves for the six extended crystals and all 23 X23 molecular crystals; decomposition of the second- and fourth-order contributions to the MP2SS direct-term correction; and additional results on $\vq$-cut convergence, log--log errors, and the effect of correcting the orbital energies with ExxSS on the MP2SS correction. A separate ZIP archive contains CSV datasets for the exploratory, extended crystal, and molecular crystal calculations, including the plotted energies, MP2SS corrections, $\vk$-point meshes, and walltimes.

\section*{Data and Software Availability}
The data underlying this study are available in the published article and in the Supporting Information, which includes the data used to plot the finite-size error curves in CSV format. The MP2SS code is also available as an installable Python package at \url{https://github.com/stephen-quiton/fsec}, along with example calculations.

\printbibliography %

\newpage

\rule{0.05in}{1.75in}%
\begin{minipage}[b][1.75in]{3.25in}
  \sffamily
  \frenchspacing
  \centering
  \includegraphics[width=\linewidth]{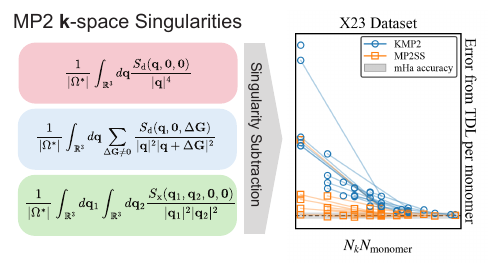}
\end{minipage}%
\rule{0.05in}{1.75in}



\section*{Supplementary material}

# Supporting Information: Reduction of finite-size effects for second-order Møller–Plesset perturbation theory with singularity subtraction

Stephen Jon Quiton[1], Juan D. F. Pottecher[1], Martin Head-Gordon*[1], and Lin Lin*[2,3]

[1]Department of Chemistry, University of California, Berkeley, California 94720, United States
[2]Department of Mathematics, University of California, Berkeley, California 94720, United States
[3]Applied Mathematics and Computational Research Division, Lawrence Berkeley National Laboratory, Berkeley, California 94720, United States

*Email: mhg@cchem.berkeley.edu, linlin@math.berkeley.edu

## S1 Basis Set Dependence of the Structure Factor

For a given crystal, the MP2 correlation energy structure factor is heavily dependent on the representation of not only the occupied bands (as is the case in HF exact exchange), but also the virtual bands. Specifically, the exchange folded structure factor and the folded second-order direct contribution both vanish as $\mathcal{O}(|\mathbf{q}|^2)$ at the origin, whereas the $\Delta\mathbf{G} = \mathbf{0}$ component of the unfolded direct structure factor vanishes as $\mathcal{O}(|\mathbf{q}|^4)$. Away from the origin, the detailed shape of the structure factor depends on which excitations are included in computing the correlation energy.

Consider the 3D-periodic isotropic Gaussian effective potential described in ref [S1], where there is only one occupied band and three virtual bands. Because the effective potential is a deep isotropic Gaussian, we can derive an approximate expression for the direct structure factor by assuming that the orbitals are relatively localized and resemble solutions of the 3D quantum harmonic oscillator in a large unit cell, so that overlap between orbitals from different cells is neglected. Specifically, we take $\psi_{1s}(\mathbf{r}) \propto e^{-\alpha r^2}$ and $\psi_{2p_x}(\mathbf{r}) \propto x e^{-\alpha r^2}$, with analogous expressions for $y$ and $z$. The Fourier transform of the MO product is then

$$\hat{\rho}_{1s2p_x}(\mathbf{G}) = \int_{\mathbb{R}^3} d\mathbf{r}\, \psi^*_{1s}(\mathbf{r})\psi_{2p_x}(\mathbf{r}) e^{i\mathbf{G}\cdot\mathbf{r}} \propto iG_x e^{-\beta_1 G^2}, \tag{1}$$

where $\beta_1 > 0$ is proportional to $1/\alpha$. Then, using the example $i = j = 1s$ and $a = b = 2p_x$,

$$r_{1s1s2p_x2p_x}(\mathbf{G}) = \hat{\rho}_{1s2p_x}(\mathbf{G})\hat{\rho}_{1s2p_x}(-\mathbf{G}) \propto G_x^2 e^{-\beta_2 G^2}, \tag{2}$$

for some $\beta_2 > 0$. Summing over the three degenerate $2p$-like virtual excitations produces the isotropic combination $\sum_{\mu,\nu=\{x,y,z\}} G_\mu(G_\nu + \Delta G_\nu) = \big(\mathbf{G}\cdot(\mathbf{G}+\Delta\mathbf{G})\big)^2$. Thus, up to a $\Delta\mathbf{G}$-dependent

prefactor that can be absorbed into the fitted amplitude, the structure factor at fixed $\mathbf{\Delta G}$ has the form

$$\tilde{S}(\mathbf{G}, \mathbf{\Delta G}) = \sum_{ijab} r_{ijab}(\mathbf{G}) \left(r_{ijab}(\mathbf{G} + \mathbf{\Delta G})\right)^* \tag{3}$$

$$\propto \left(\mathbf{G} \cdot (\mathbf{G} + \mathbf{\Delta G})\right)^2 \exp\left(-\beta_3 \left|\mathbf{G} + \frac{\mathbf{\Delta G}}{2}\right|^2\right). \tag{4}$$

Here $\beta_3 > 0$, and we have absorbed the constant factor $\exp(-\beta_3|\mathbf{\Delta G}|^2/4)$ into the overall amplitude because the fit is carried out separately for each chosen value of $\mathbf{\Delta G}$. This motivates the auxiliary function

$$f(\mathbf{q}, \mathbf{\Delta G}) = c_0 \left(\mathbf{q} \cdot (\mathbf{q} + \mathbf{\Delta G})\right)^2 \exp\left(-\beta_3 \left|\mathbf{q} + \frac{\mathbf{\Delta G}}{2}\right|^2\right), \tag{5}$$

with $c_0$ and $\beta_3$ determined by least squares. Figure S1 shows representative fits for $\mathbf{\Delta G} = \mathbf{0}$ and $\mathbf{\Delta G} \neq \mathbf{0}$.

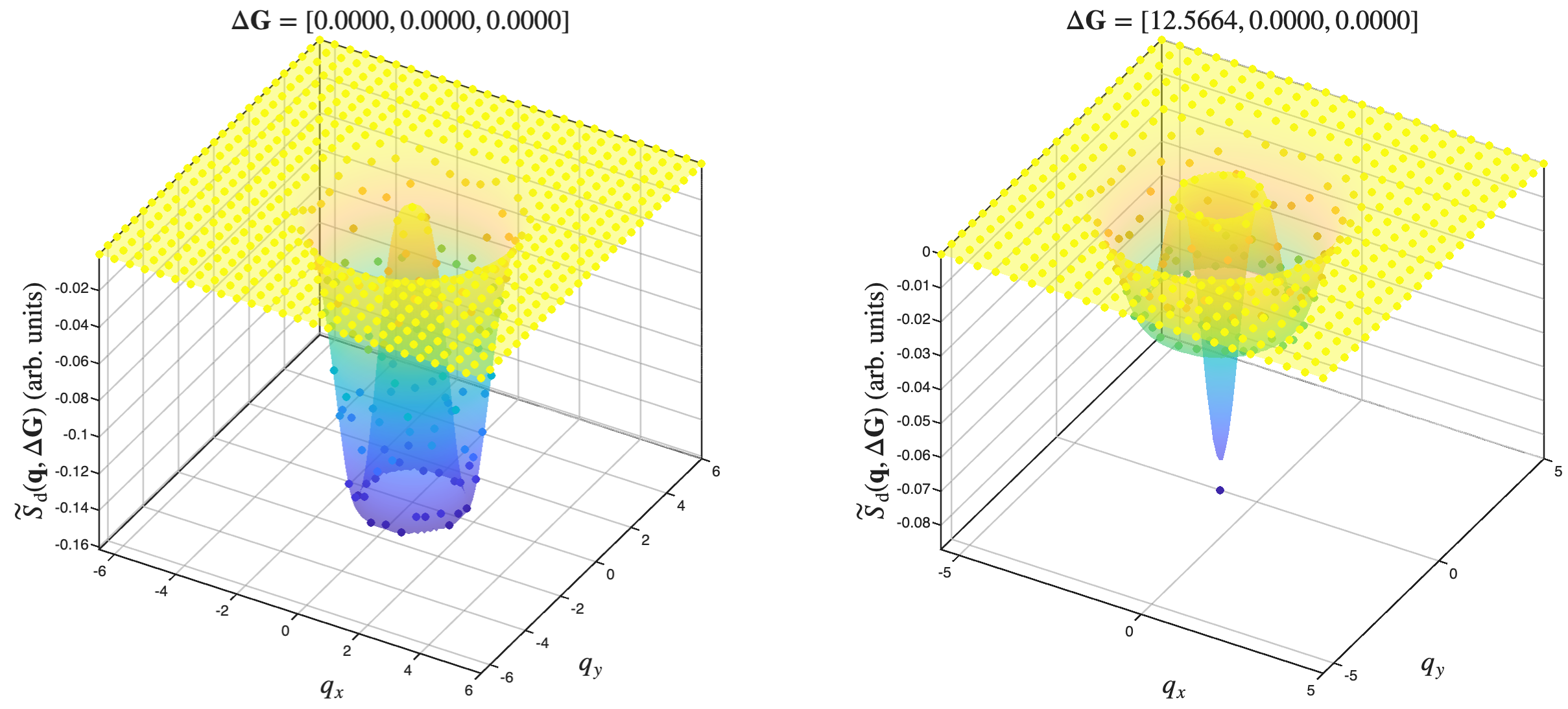


(a) (b)

Figure S1: Unfolded direct structure factor at $\mathbf{\Delta G} = \mathbf{0}$ (a) and $\mathbf{\Delta G} \neq \mathbf{0}$ (b) for the 3D-periodic Gaussian isoptropic effective potential model system described in ref [S1] at $N_{\mathbf{k}} = 2 \times 2 \times 2$. The dots are the true structure factor, and the surface is the fitted auxiliary function given in eq (5).

The derivation above is meant to be illustrative of how the shape of the structure factor away from the origin is *directly* dependent on what excitations are included in the correlation energy. The inclusion of more virtual bands will lead to more complicated structure factor expressions, further underscoring the need to reach the complete basis set limit not only in light of capturing the MP2 correlation energy but also when developing an appropriate auxiliary function for singularity subtraction.

## S2 FSE curves with unshifted staggered mesh

In the main text, all finite-size error (FSE) plots (except for diamond with the GTH-SZV basis) contained a line for Staggered Mesh that was shifted vertically such that the highest **k**-mesh result coincided with the TDL estimate. Below are all plots *without* the shift.

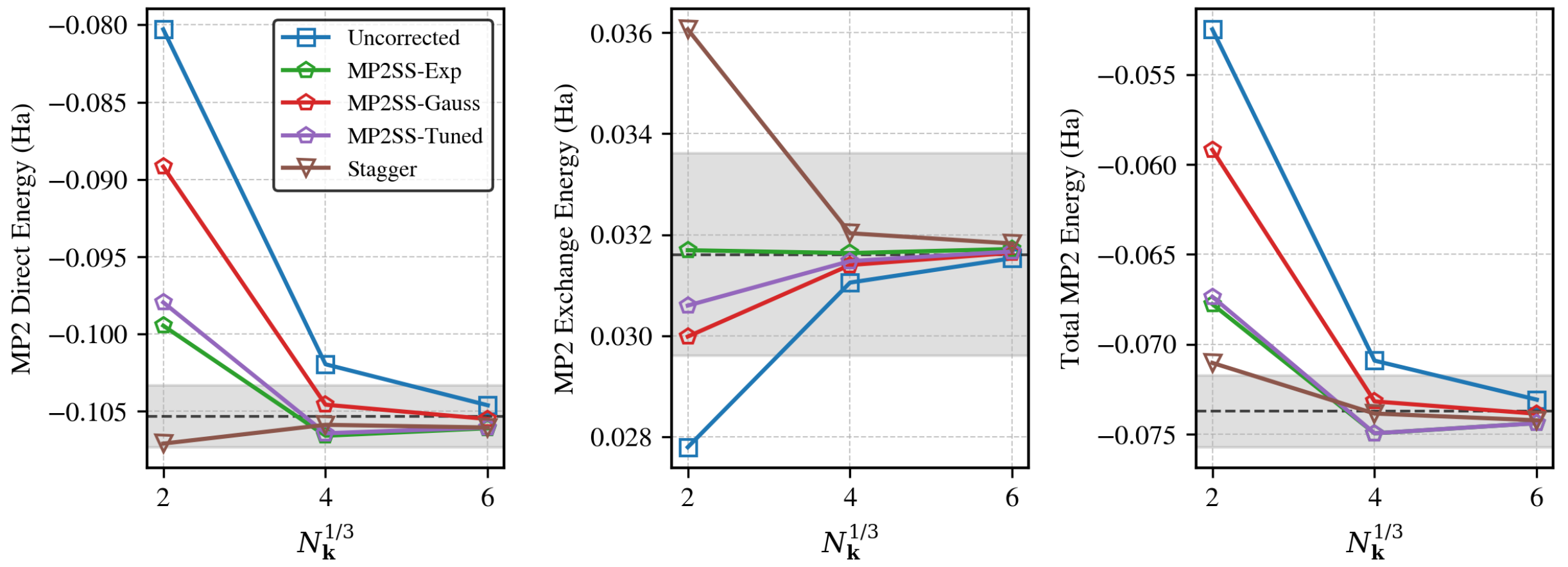


Figure S2: Finite-size effect of the direct, exchange, and total MP2 correlation energy per unit cell with the GTH-SZV basis set for silicon with different MP2SS configurations. The gray horizontal stripe represents millihartree accuracy per atom with respect to the TDL, estimated by an $N_{\mathbf{k}}^{-1}$ power-law fit with the uncorrected curve.

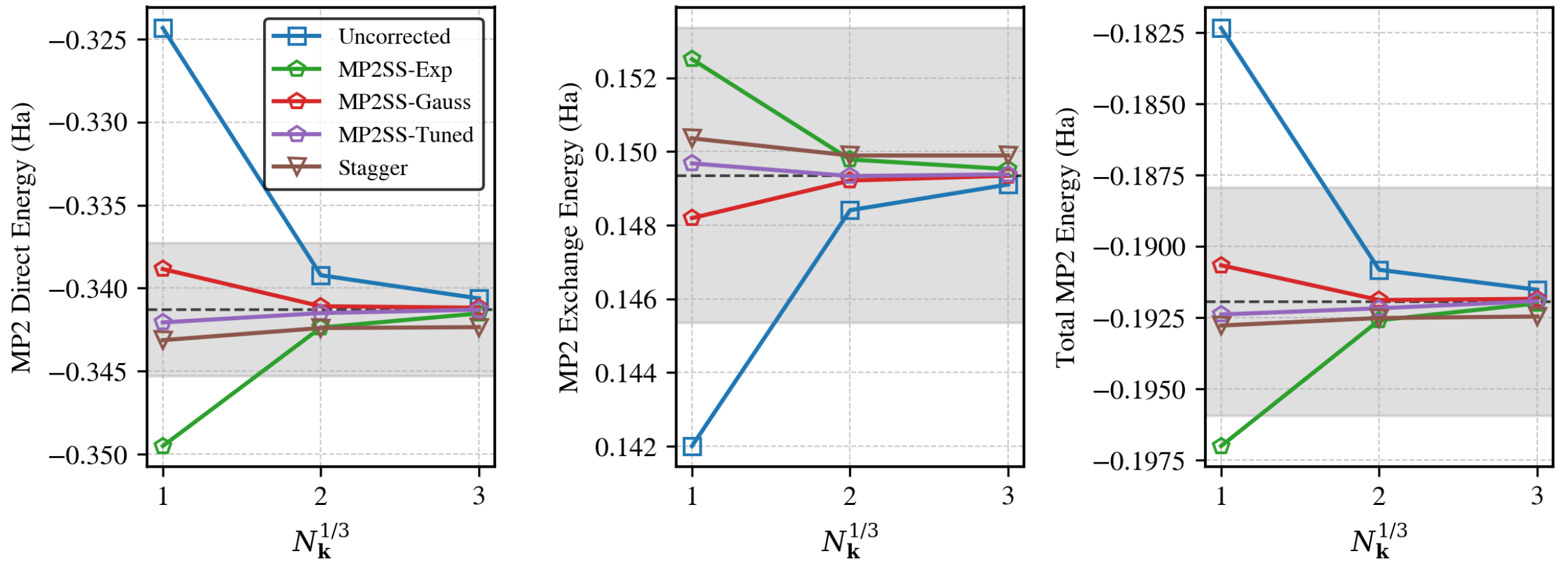


Figure S3: Finite-size effect of the direct, exchange, and total MP2 correlation energy per unit cell with the GTH-SZV basis set for ammonia with different MP2SS configurations. The gray horizontal stripe represents millihartree accuracy per monomer with respect to the TDL, estimated by an $N_{\mathbf{k}}^{-1}$ power-law fit with the uncorrected curve.

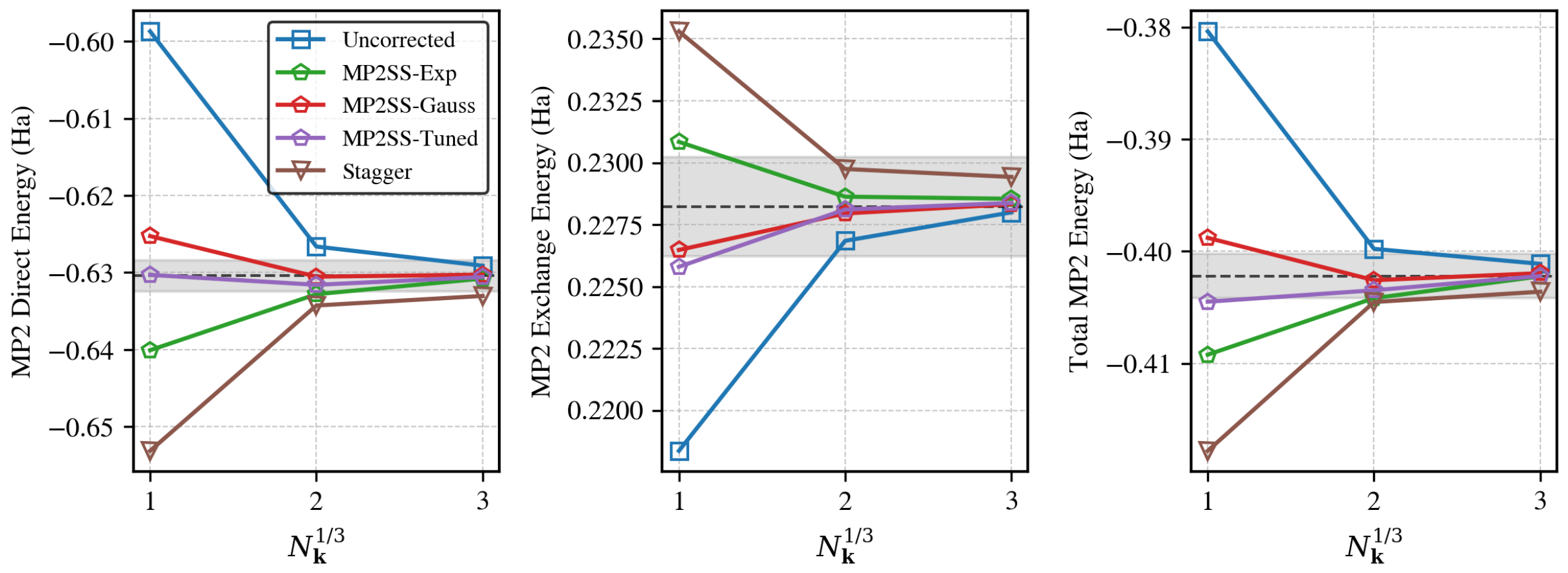


Figure S4: Finite-size effect of the direct, exchange, and total MP2 correlation energy per unit cell with the GTH-SZV basis set for urea with different MP2SS configurations. The gray horizontal stripe represents millihartree accuracy per monomer with respect to the TDL, estimated by an $N_{\mathbf{k}}^{-1}$ power-law fit with the uncorrected curve.

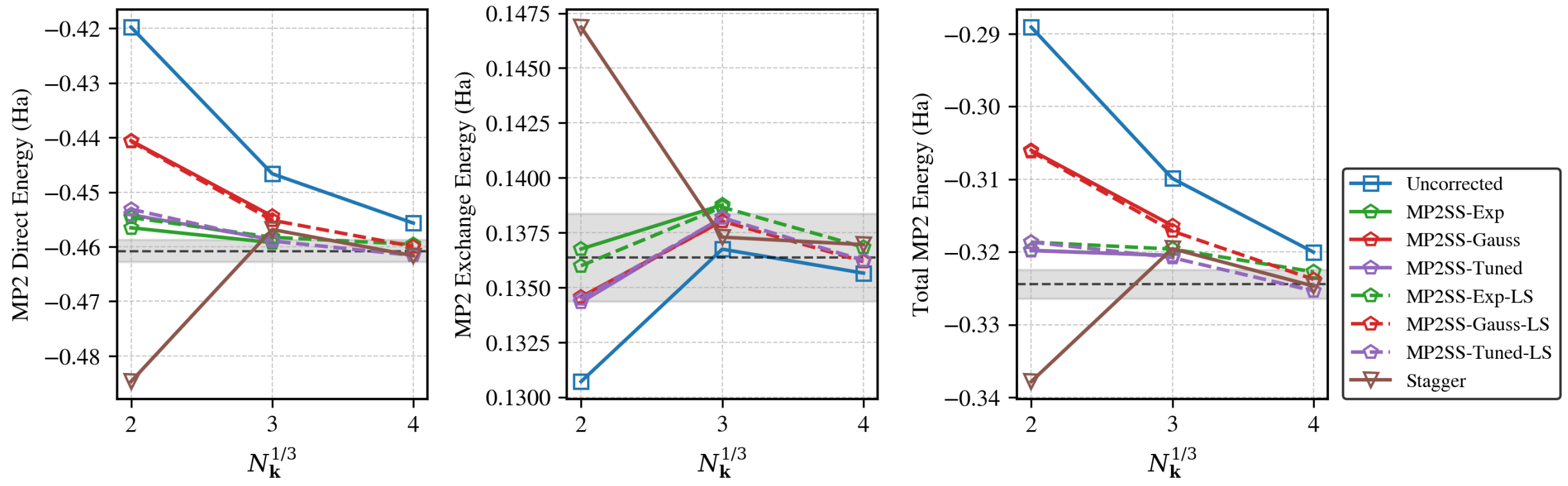


Figure S5: Finite-size effect of the direct, exchange, and total MP2 correlation energy per unit cell with the GTH-cc-pVTZ basis set for diamond with different MP2SS configurations. The gray horizontal stripe represents millihartree accuracy per atom with respect to the TDL, estimated by an $N_{\mathbf{k}}^{-1}$ power-law fit with the uncorrected curve. The data point at $N_{\mathbf{k}} = 3^3$ is not included to obtain the TDL estimate. The dashed curves represent MP2SS corrections when the auxiliary function is fit to line-sampled (LS) structure factor data. See text for details.

# S3 Second and Fourth Order Contributions to the MP2SS Direct Term Correction

Below are plots that decompose the quadratic and quartic contributions to the MP2SS correction to the direct term for each MP2SS configuration in the text.

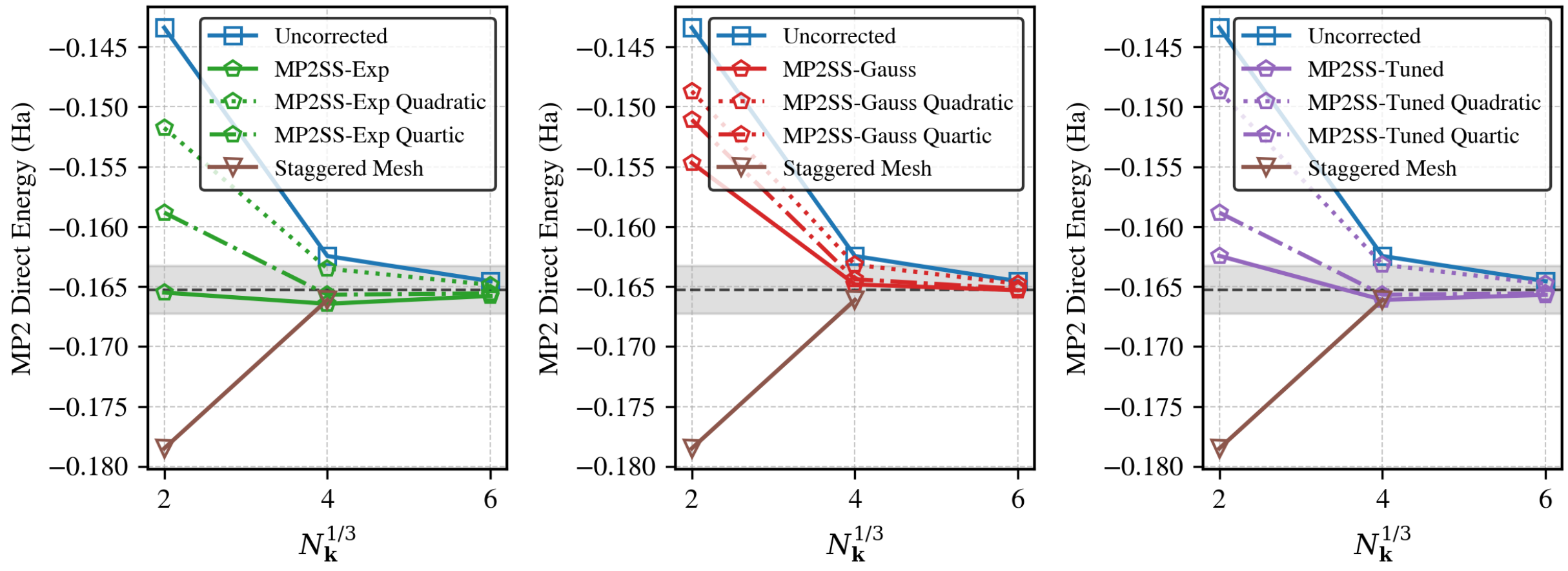


Figure S6: FSE curves for the direct term of MP2 for diamond with the GTH-SZV basis set. Each plot is a different MP2SS configuration. Included are the total, quadratic-only, and quartic-only MP2SS corrections to the direct term. The gray horizontal stripe represents millihartree accuracy per atom with respect to the TDL, estimated by an $N_{\mathbf{k}}^{-1}$ power-law fit with the uncorrected curve.

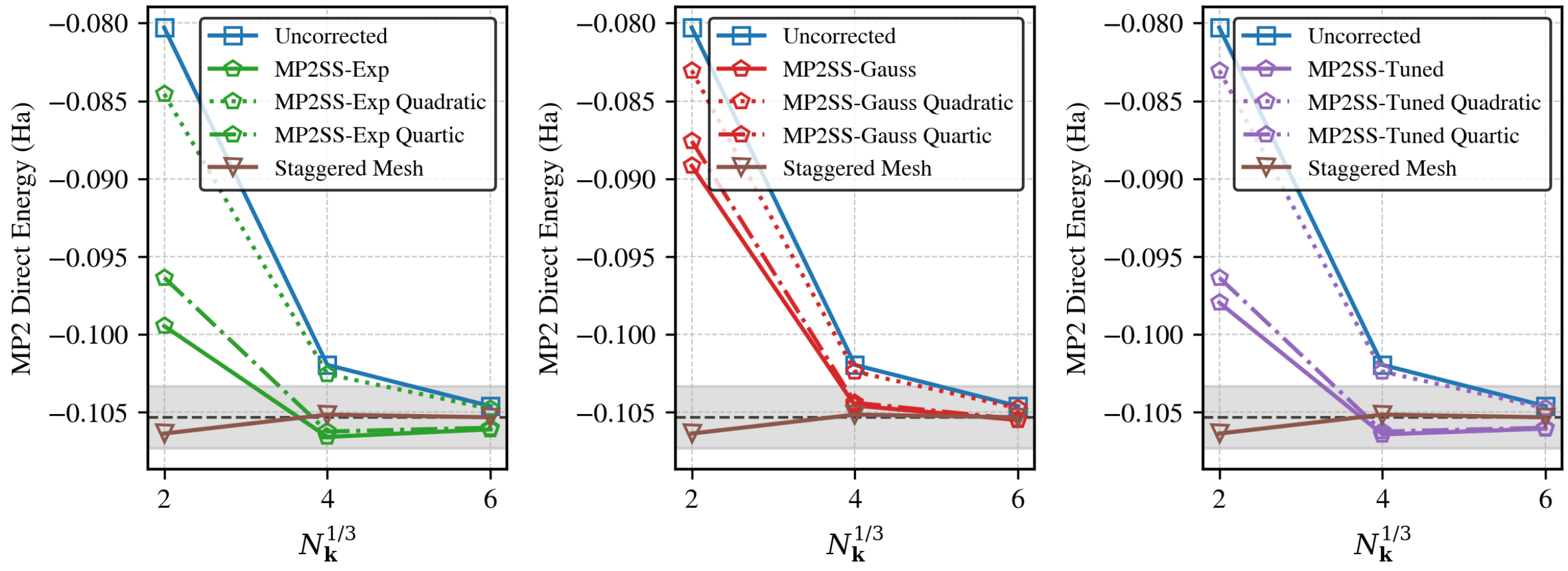


Figure S7: FSE curves for the direct term of MP2 for silicon with the GTH-SZV basis set. Each plot is a different MP2SS configuration. Included are the total, quadratic-only, and quartic-only MP2SS corrections to the direct term. The gray horizontal stripe represents millihartree accuracy per atom with respect to the TDL, estimated by an $N_{\mathbf{k}}^{-1}$ power-law fit with the uncorrected curve.

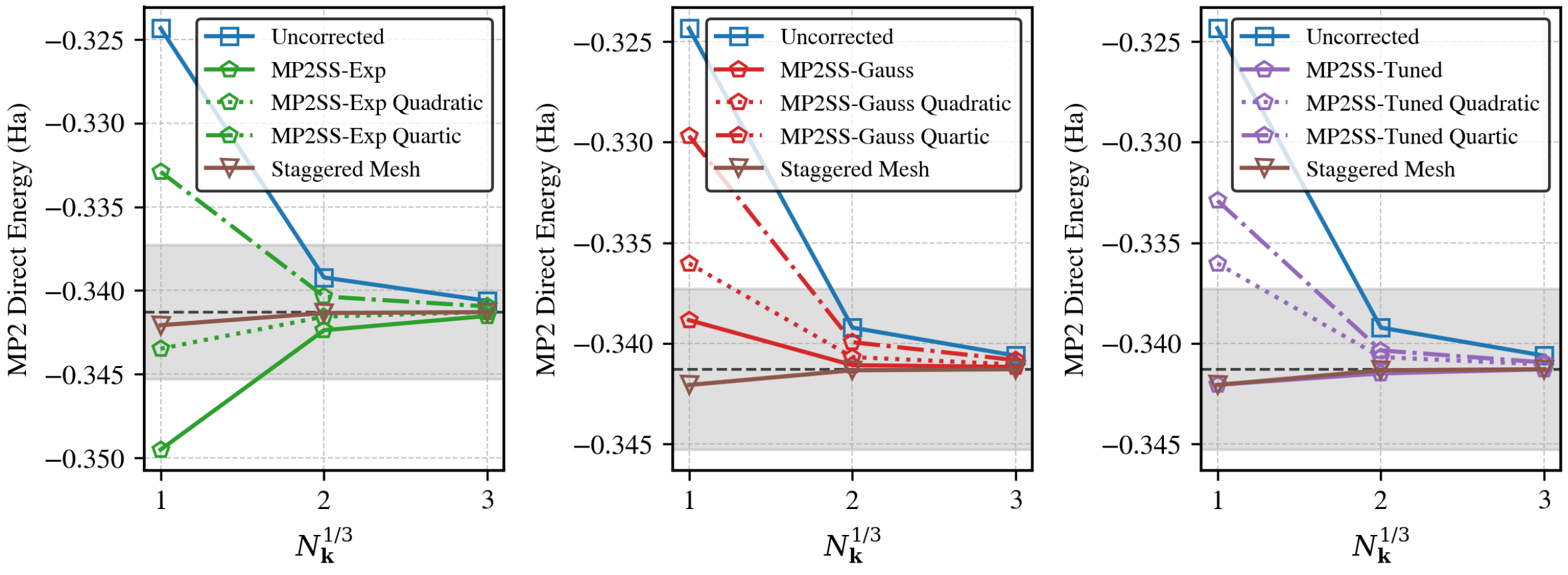


Figure S8: FSE curves for the direct term of MP2 for ammonia with the GTH-SZV basis set. Each plot is a different MP2SS configuration. Included are the total, quadratic-only, and quartic-only MP2SS corrections to the direct term. The gray horizontal stripe represents millihartree accuracy per atom with respect to the TDL, estimated by an $N_{\mathbf{k}}^{-1}$ power-law fit with the uncorrected curve.

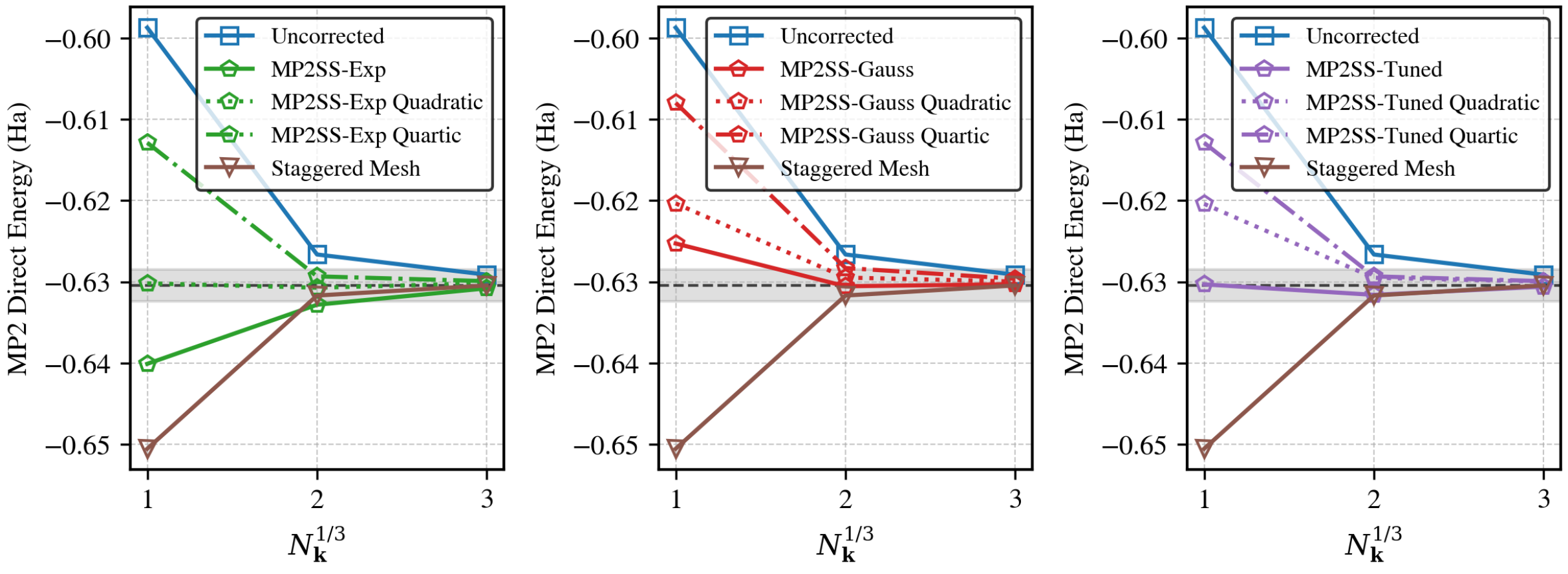


Figure S9: FSE curves for the direct term of MP2 for urea with the GTH-SZV basis set. Each plot is a different MP2SS configuration. Included are the total, quadratic-only, and quartic-only MP2SS corrections to the direct term. The gray horizontal stripe represents millihartree accuracy per atom with respect to the TDL, estimated by an $N_{\mathbf{k}}^{-1}$ power-law fit with the uncorrected curve.

## S4 Preliminary Walltime Benchmarks

While the main focus of this work is to demonstrate singularity subtraction's promise in correcting FSEs in the correlation energy, we can also show that MP2SS already reduces the computational resources necessary to achieve convergence towards the TDL. To briefly illustrate this, we benchmark the time it takes for uncorrected MP2 and the staggered mesh method to reach the same level of accuracy towards TDL as MP2SS-Tuned for diamond and ammonia. The results are shown in figure S10. Just as in ExxSS, the current main bottleneck of the calculation is the construction of the structure factors. This portion, however, may be easily accelerated in the SS scheme because the only role of the structure factor samples is to obtain a good fit for the auxiliary function. Our preliminary results with MP2SS-LS for diamond using a triple zeta quality basis shown in figure 9 in the main text suggest that not all points need to be sampled in order to achieve significantly improved convergence toward the TDL. We reserve further optimization of the MP2SS procedure for a future work.

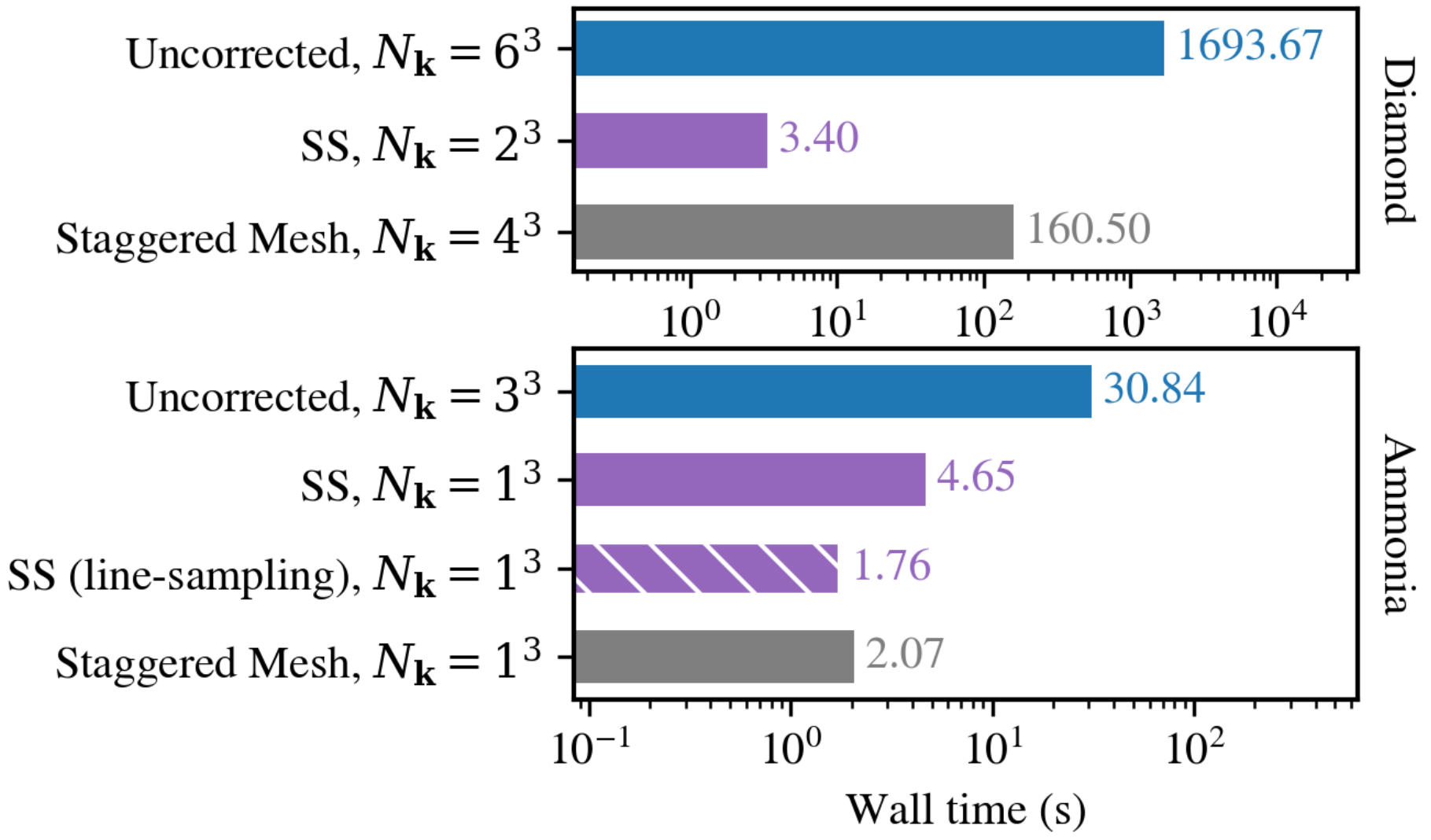


Figure S10: Walltime comparisons between uncorrected MP2, singularity subtraction-corrected MP2, and staggered-mesh MP2 for diamond and ammonia with the GTH-SZV basis set, corresponding to figures 5 and 7 respectively in the main text. The **k**-meshes for uncorrected and staggered mesh were chosen based on what achieves similar or better accuracy as SS at the sparsest **k**-mesh. All walltimes were achieved with an Intel Xeon Skylake 6130 using 8 threads.

## S5 Additional Results

Below are data and plots of additional numerical experiments to support the main text.

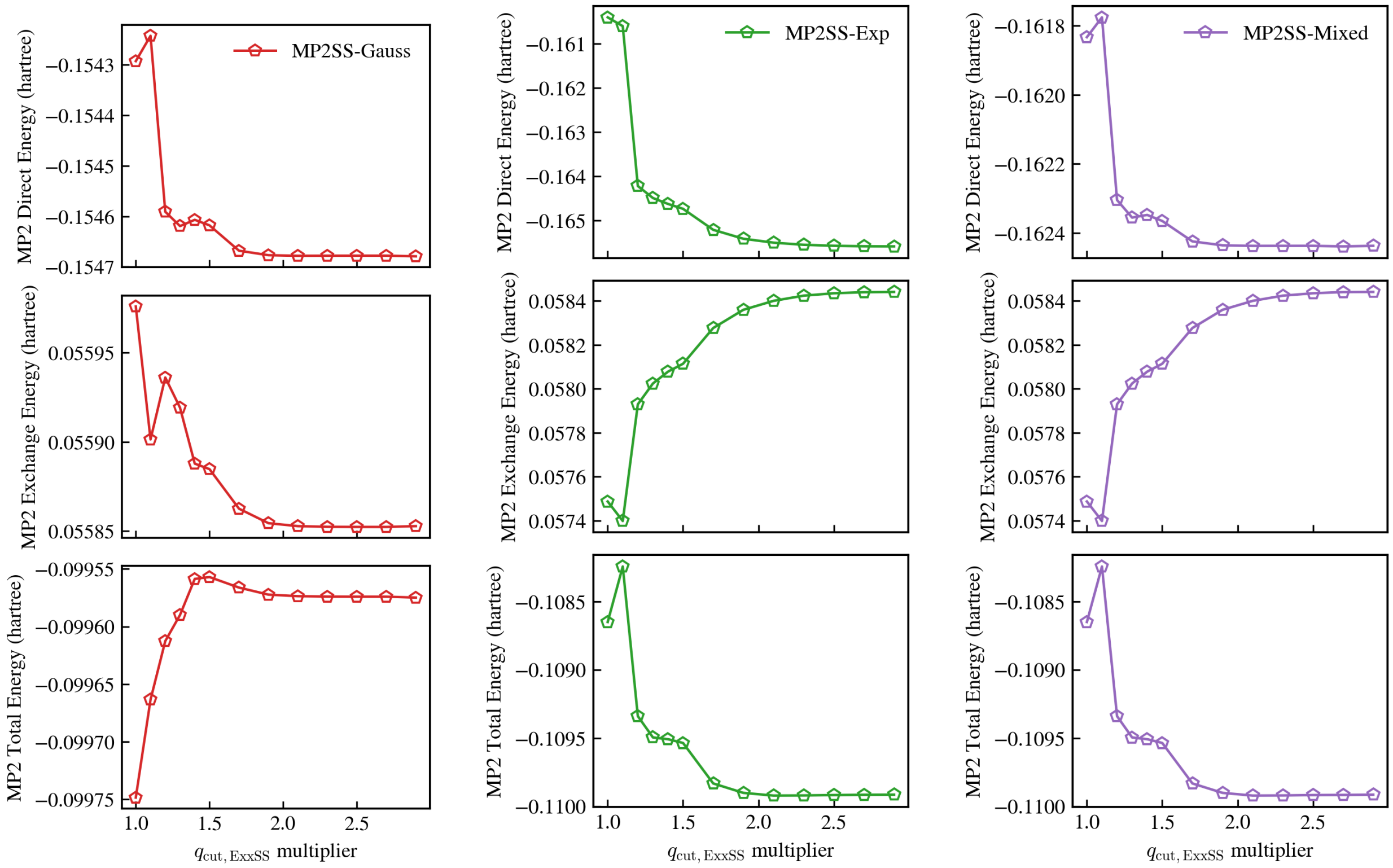


(a) (b) (c)

Figure S11: Effect of the cutoff radius for structure factor sampling on the corrected MP2 energy components for diamond at GTH-SZV. The x-axis represents the multiplier of the radius set by ExxSS [S2], which is 1.796 bohr$^{-1}$. The default multiplier for all results in the main text is set to 2.0.

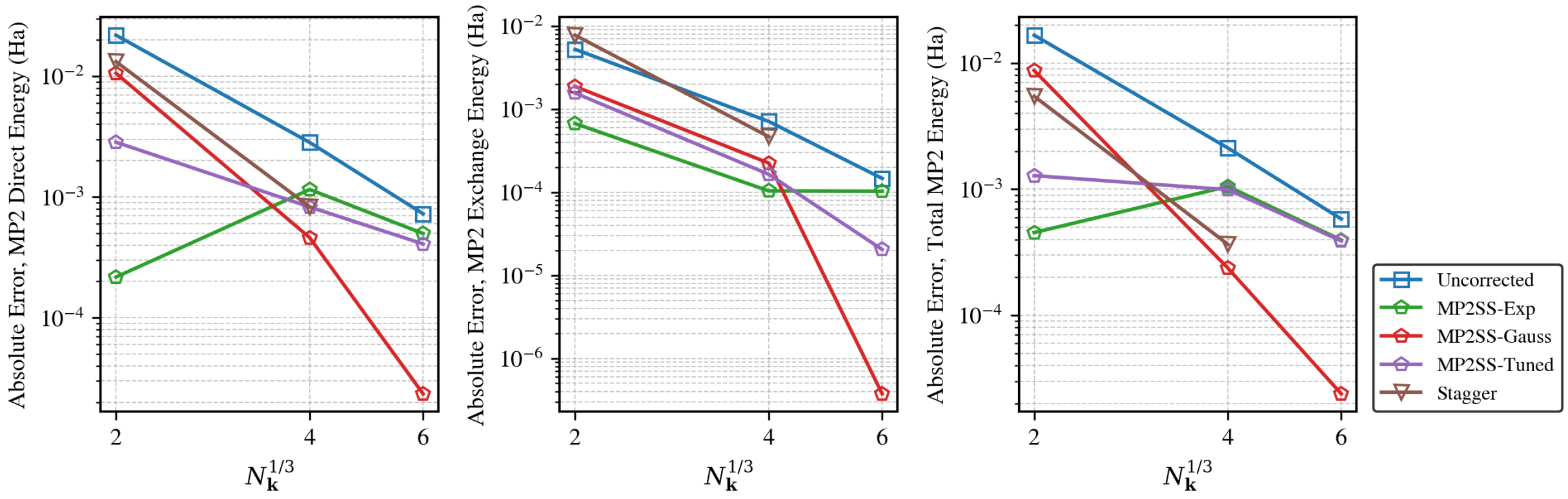


Figure S12: Log-log absolute error plot of the MP2SS configurations vs. the uncorrected energy and the staggered mesh method for diamond at GTH-SZV. The reference is estimated from an $N_{\mathbf{k}}^{-1}$ power-law fit to the uncorrected energies for each term.

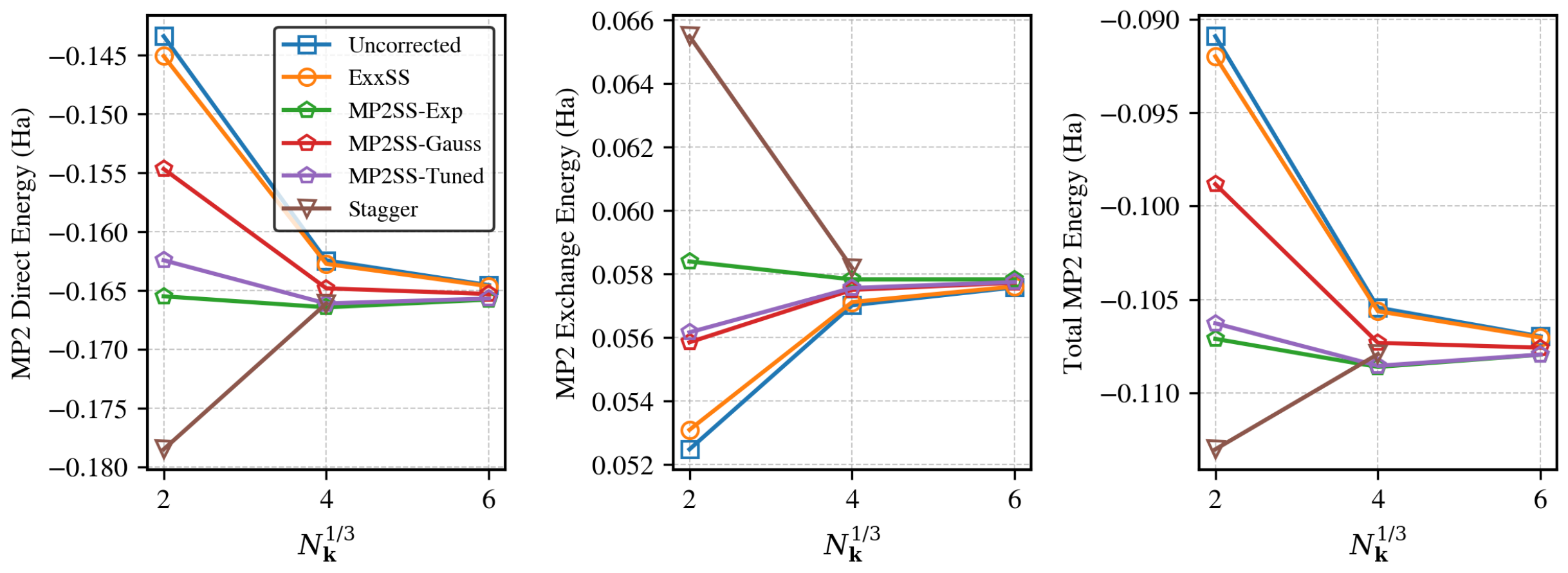


Figure S13: FSE convergence plot of diamond using the GTH-SZV basis set. Shown as an orange line is the FSE correction provided *only* by ExxSS and not MP2SS. ExxSS shifts the occupied orbital energies to correct the FSE in the HF exact exchange energy.

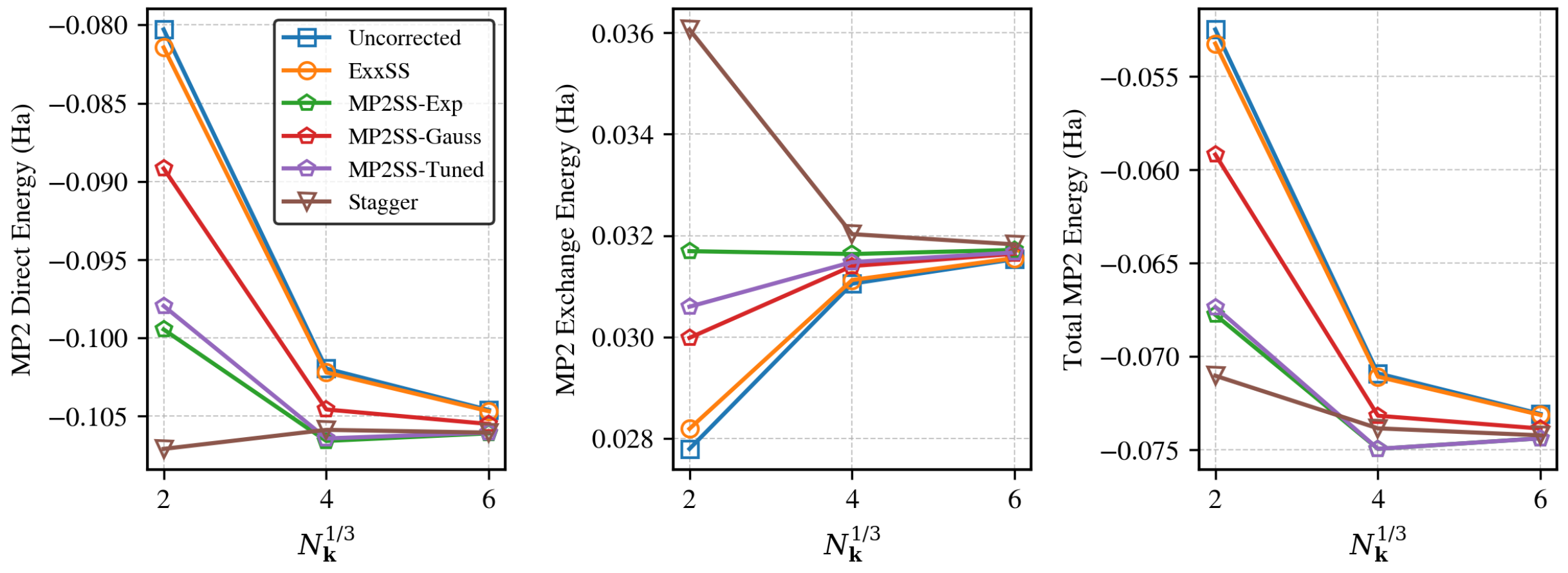


Figure S14: FSE convergence plot of silicon using the GTH-SZV basis set. Shown as an orange line is the FSE correction provided *only* by ExxSS and not MP2SS. ExxSS shifts the occupied orbital energies to correct the FSE in the HF exact exchange energy.



\end{document}